\documentclass[%
 onecolumn,
 amsmath,amssymb,
 aps, showkeys
]{revtex4-2}

\usepackage{graphicx}
\usepackage{dcolumn}
\usepackage{bm}
\usepackage{hyperref}
 \usepackage{booktabs}
 \usepackage{comment}
 \usepackage{orcidlink}
\usepackage{multirow}
\usepackage{fontawesome5} 
\usepackage{hyperref}    

\begin{document}


\title{Comparing measures of the Hubble and BAO tensions in \texorpdfstring{$\Lambda$}{L}CDM and possible solutions in \texorpdfstring{$f(Q)$}{f(Q)} gravity}

\newcommand{\orcidauthorA}{0000-0001-9738-7704 } 
\newcommand{\orcidauthorB}{0000-0003-0685-9791} 
\newcommand{\orcidauthorC}{0000-0002-4123-7325}

\author{Jos{\'e} Antonio N{\'a}jera \orcidlink{0000-0001-9738-7704}}
\email{antonio.najera@port.ac.uk}
\author{Indranil Banik \orcidlink{0000-0002-4123-7325}}
\email{indranil.banik@port.ac.uk}
\author{Harry Desmond \orcidlink{0000-0003-0685-9791}}
 \email{harry.desmond@port.ac.uk}
\affiliation{Institute of Cosmology \& Gravitation, University of Portsmouth, Dennis Sciama Building, Portsmouth, PO1 3FX, UK}

\author{Vasileios Kalaitzidis}
\email{vasileioskalaitzidis02@gmail.com}
\affiliation{Scottish Universities Physics Alliance, University of Saint Andrews, North Haugh, Saint Andrews, Fife KY16 9SS, UK}

\date{\today}

\begin{abstract}
We test whether $f(Q)$ symmetric teleparallel gravity theories can solve the Hubble tension consistently with DESI~DR2 BAO. We consider three $f(Q)$ functional forms: logarithmic, exponential, and hyperbolic tangent. We extend these models by allowing a cosmological constant, and compare to phenomenological models with a flexible exponential, hyperbolic secant, and polynomial decay addition to the standard $\Lambda$CDM $H(z)$. We test these models against DESI~DR2 BAO, CMB (\emph{Planck}~2018 + SPT-3G + ACT~DR6), local $H_0$, and Cosmic Chronometer data. The logarithmic and hyperbolic tangent $f(Q)$ models do not provide an adequate solution, but the exponential model does. Furthermore, it slightly reduces the $(\Omega_m, H_0 r_d)$ parameter space tension between CMB and BAO datasets to $2.56\sigma$, down from $2.65\sigma$ for $\Lambda$CDM. Although $\Lambda$CDM faces only $1.66\sigma$ tension in DESI data space, the $1\sigma$ higher tension in parameter space suggests a real anomaly. The models assisted by the cosmological constant perform slightly better still, at the cost of undermined theoretical motivation. They also perform poorly once local $H_0$ measurements are included. The phenomenological models fit all data reasonably well, yet the best-fitting models predict isotropically averaged BAO distances exceeding the DESI~DR2 measurements at all redshifts. This highlights the difficulties of finding a theoretically motivated solution to the Hubble tension while remaining consistent with BAO data.
\end{abstract}

\maketitle


\section{Introduction}




If General Relativity \citep[GR;][]{Einstein_1915} is valid on cosmological scales, the matter content of the Universe must be dominated by non-baryonic cold dark matter (CDM) particles outside the standard model of particle physics, while the matter-energy content must be dominated by a cosmological constant ($\Lambda$), leading to the well-known $\Lambda$CDM paradigm \citep{Efstathiou_1990, Ostriker_1995}. Its free parameters can be calibrated using the observed anisotropies in the power spectrum of the cosmic microwave background (CMB), relic radiation from redshift $z = 1090$ \citep{Planck_2020}. One of the model parameters is $H_0$, the present value of the Hubble parameter $H \equiv \dot{a}/a$, where $a$ is the cosmic scale factor normalised to unity today, 0 subscripts denote present values, and an overdot denotes a time derivative. The $\Lambda$CDM prediction for $H_0$ based on the CMB anisotropies measured by the \emph{Planck} satellite, the Atacama Cosmology Telescope (ACT), and the South Pole Telescope (SPT) is $67.24 \pm 0.35$~km/s/Mpc \citep{Planck_2020, Calabrese_2025, Camphuis_2025}. This can be tested by measuring $z' \equiv dz/dr$, the local gradient of redshift with respect to distance. In a homogeneously expanding universe, we expect that $cz' = H_0$ (section~1 of \citep{Mazurenko_2025}). However, local measurements of $cz'$ by several different teams using a wide variety of techniques, instruments, and tracers consistently indicate a value around 9\% above that expected in $\Lambda$CDM \citep[][and references therein]{H0DN_2025, Valentino_2025}.

$\Lambda$CDM also faces another tension from the observed angular scale and redshift depth of the baryon acoustic oscillation (BAO) ruler. This is a feature in the large-scale correlation function of galaxies that was imprinted at early times, essentially corresponding to the first angular peak in the CMB power spectrum \citep{Eisenstein_1998}. The comoving length $r_d$ of the BAO ruler should have remained constant for $z \lesssim 1060$. Starting with the seminal works of \cite{Cole_2005} and \cite{Eisenstein_2005}, the BAO feature has been detected at increasingly higher precision, most recently by the Dark Energy Spectroscopic Instrument (DESI) Collaboration in their Data Release~2 (DR2) \citep{DESI_2025}. DESI DR2 is particularly important since it provides crucial constraints on the expansion history at intermediate redshifts. The measurements are in $\approx 3\sigma$ tension with $\Lambda$CDM expectations \citep{Calabrese_2025}. However, a look at all available BAO measurements over the last 20 years shows that the actual level of tension is more like $3.8\sigma$, highlighting that the BAO `anomaly' has to be taken seriously (table~2 of \citep{Banik_2025_BAO}).

These difficulties with $\Lambda$CDM might suggest a problem with our understanding of the $\mathcal{O} \left( 10^{-5} \right)$ CMB anisotropies. However, we can infer $H_0$ using only the CMB monopole temperature in combination with the primordial light element abundances, relying on the well-established theory of Big Bang nucleosynthesis \citep[BBN;][]{Cyburt_2016}. This BAO + BBN route to $H_0$ gives $68.51 \pm 0.58$~km/s/Mpc \citep{DESI_2025}, far below local measurements of $cz'$. Thus, problems with measurements of the subtle CMB anisotropies cannot account for the Hubble tension.

The above argument relies on $\Lambda$CDM applying prior to recombination, but if there are actually significant deviations from it, then we may be able to fit the CMB anisotropies and BBN with a higher $H_0$ consistent with the local $cz'$. There have been several proposals along these lines, for instance using primordial magnetic fields to increase the redshift of recombination \citep{Mirpoorian_2025, Jedamzik_2026}. However, there are at least seven serious difficulties faced by such early-time solutions to the Hubble tension \citep{Vagnozzi_2023}. One such difficulty is that the age of the Universe would be reduced by a higher expansion rate than the \emph{Planck} cosmology for nearly the entirety of cosmic history \citep{Vagnozzi_2022}. However, this prediction is in significant tension with observations of the oldest Galactic stars and globular clusters \citep{Valcin_2021, Cimatti_2023, Xiang_2022, Souza_2024, Lundkvist_2025, Shariat_2025, Tomasetti_2025, Valcin_2025, Xiang_2025}, which are more in line with the \emph{Planck} $H_0$ given the constraint on the matter density parameter $\Omega_{\mathrm{M}}$ from uncalibrated BAO data \citep{Banik_2025_cosmology}. Another issue is that the CMB contains a wealth of detail beyond merely the angular scale of the first acoustic peak. This makes it challenging to understand the excellent fit to the CMB power spectrum in $\Lambda$CDM if there were indeed substantial deviations from it in the pre-recombination Universe \citep{Calabrese_2025, Camphuis_2025}. Moreover, these deviations would need to be tuned such that the horizon scale at matter-radiation equality ($z \approx 3400$) agrees very well with the $\Lambda$CDM prediction calibrated using the sound horizon at recombination \citep{Zaborowski_2025}. Another issue is that the BAO anomaly is not present at high $z$ but grows as we get to $z \lesssim 1$. This trend cannot be understood through a simple recalibration of $r_d$, which would uniformly rescale the BAO angular scale and redshift depth at any $z$ by the same factor. Several other studies have also reported that the $H_0$ inferred from data at some redshift declines with $z$ \citep{Jia_2023, Jia_2025a, Lopez_2025}. This descending trend is very indicative of departures from $\Lambda$CDM in the late Universe. If we assume standard physics at early times and allow a flexible expansion rate history, it is possible to predict that $H_0$ should have a value similar to the measured local $cz'$, even when this is not used as a constraint \citep{Jia_2025b}. This connects early and late Universe measurements using a departure from the standard Friedmann equations only at $z \lesssim 0.5$.

These difficulties faced by $\Lambda$CDM and modifications to it at early times motivate us to consider other theories with a different late-time expansion history. In fact, there is robust evidence for considering alternatives to $\Lambda$CDM from the $\approx 5\sigma$ C.L. significance supporting a quintessence-phantom transition at $z \approx 0.5$ \citep{Scherer_2025}. This analysis based on CMB, BAO, and uncalibrated Type~Ia supernovae (SNIa) constitutes one of the strongest results supporting the need for dynamical dark energy or modified gravity, though the specific model considered by those authors cannot solve the Hubble tension. It has also been suggested that interactions between the dark matter and dark energy sectors could alleviate the Hubble and BAO tensions \citep{Zhang_2025}. Sticking to $\Lambda$CDM behaviour in the early Universe would preserve the excellent fit to the CMB anisotropies in $\Lambda$CDM, provided the comoving matter density and the comoving distance to recombination remain the same as in the \emph{Planck} cosmology. This still leaves significant flexibility in the late-time expansion history, possibly allowing us to solve the Hubble and BAO tensions. To modify the expansion history, we need to alter either the matter-energy content of the universe or the law of gravity. In the former approach, we can introduce dynamical dark energy, where we let the dark energy Equation of State (EoS) change with redshift. However, this proposal does not explain what dark energy is, so it just provides a phenomenological approach without theoretical grounds. The latter approach is through modified theories of gravity. In this context, the late-time accelerated expansion of the Universe is described via its geometry. This provides a compelling explanation for the nature of dark energy.

In this paper, we focus on the latter approach, i.e., the modified gravity formalism. Specifically, we work in the symmetric teleparallel framework, where curvature and torsion are zero and gravity is driven by non-metricity (the covariant derivative of the metric) \citep{Nester_1998}. In this background, it is possible to build the symmetric teleparallel equivalent of General Relativity \citep[(STEGR);][]{Beltran_2019} given by the Lagrangian $\mathcal{L} = Q/(8\pi G)$. Similarly to Riemannian geometry where we can promote $R \to f(R)$ \citep{DeFelice_2010}, we can generalize the Lagrangian to a general function of the non-metricity scalar $Q \to f(Q)$ \citep{Jimenez_2018}. This gives us a set of theories called $f(Q)$ gravity.

There have been numerous recent studies on $f(Q)$ gravity. The background cosmology and first order perturbations were studied in \citep{Jimenez_2020}. The theory was tested with redshift space distortions (RSD) in \citep{Barros_2020}. There are also studies in cosmography \citep{Mandal_2020_cosmography}, energy conditions \citep{Mandal_2020_energy}, and quantum cosmology \citep{Dimakis_2021}. The most general connection for the symmetric teleparallel framework was proposed in \citep{Ambrosio_2022, Hohmann_2021}. Different possible functional forms for $f(Q)$ have been considered which describe an accelerated expansion of the Universe at late times and are possible alternatives to $\Lambda$CDM, including power-law \citep{Mandal_2023}, exponential \citep{Fotios_2021}, logarithmic \citep{Najera_2023}, and hyperbolic tangent \citep{Fotios_2023}. The propagation of gravitational waves in $f(Q)$ gravity has been considered in \citep{DAgostino_2022, Najera_2023, Capozziello_2024_gauge, Capozziello_2024_geodesic}. Solar system tests for $f(Q)$ gravity were done in \citep{Wang_2024}. Some studies have proposed solutions to the Hubble tension within $f(Q)$ gravity \citep{Sakr_2024, Boiza_2025}. A recent study found that the power-law hyperbolic tangent $f(Q)$ model alleviates the Hubble tension using DESI + RSD + Cosmic Chronometers (CC) + Standard Sirens + CMB data \citep{Kavya_2025}. Compared to their study, we consider a wider range of $f(Q)$ functional forms.

Since $f(Q)$ models have achieved some observational success as previously described, it is natural to consider whether they can solve recent cosmological tensions. In the present paper, we focus on testing whether $f(Q)$ gravity models can solve the Hubble tension and at the same time achieve BAO consistency. We work with DESI~DR2 data \citep{DESI_2025}, \emph{Planck}~2018 constraints \citep{Planck_2020}, local $H_0$ measurements like SH0ES \citep{Riess_2022_comprehensive}, and Cosmic Chronometers (or Clocks). These data help to see whether $f(Q)$ gravity is capable of solving the tensions in cosmology. We focus on three models: logarithmic \cite{Najera_2023}, exponential \cite{Fotios_2021}, and hyperbolic-tangent \cite{Fotios_2023}. These models have been found to predict an accelerated expansion of the Universe without dark energy, so we see if they can also solve the current problems in cosmology. To complement this, we also explore flexible phenomenological models which can change the expansion history $H(z)$. The objective of this is to determine whether a background solution to the Hubble and BAO tensions is feasible, either in the more theoretically motivated $f(Q)$ models or in the phenomenologically motivated models. In all cases, we carefully consider the extent to which models with parameters calibrated using non-BAO datasets can predict the BAO data, either in DESI data space or in the $(\Omega_m, H_0 r_d)$ parameter space. This provides a much more rigorous assessment of whether these models can solve the BAO anomaly.

The structure of this paper is as follows: In section~\ref{sec:fQGravityTheory}, we present the theoretical formalism for $f(Q)$ gravity. Section~\ref{sec:fQ_models} presents the specific models we consider. We describe the observational data used to constrain the models in section~\ref{sec:ObsData}. We present our findings in section~\ref{sec:Results} and discuss their implications in section~\ref{sec:Discussion}. We conclude in section~\ref{sec:Conclusions}. In Appendix~\ref{sec:FiguresBAO}, we present some figures comparing the models to all available BAO data over the last 20 years.

\section{\texorpdfstring{$f(Q)$}{f(Q)} Gravity Theory}
\label{sec:fQGravityTheory}


Gravity is a geometrical entity that is usually built upon Riemannian Geometry, where the only non-vanishing quantity is the Riemann tensor $R^\alpha_{\;\;\beta\mu\nu}$. The standard $\Lambda$CDM cosmological model was proposed within this formalism. However, the most general type of geometry is called \textit{metric-affine geometry} consisting of a triple $(\mathcal{M}, g, \Gamma)$ \citep{Capozziello_2022, Capozziello_2023, Snapper_2014}. For this, we need to choose a manifold $\mathcal{M}$, a metric $g$, and a connection $\Gamma$. Given a connection, there is a notion of parallel transport. Its effects on vectors can be quantified by three tensors. The Riemann tensor $R^\alpha_{\;\;\beta \mu\nu}$ quantifies how a vector changes its direction when moved on an infinitesimal closed path. We now consider two infinitesimal direction vectors ($\delta v_a$, $\delta v_b$) and perform a parallel transport first along $\delta v_a$ and then along $\delta v_b$. The torsion tensor $T^\alpha_{\;\;\mu\nu}$ quantifies the difference in parallel transport from this operation and the parallel transport along $\delta v_b$ first and then $\delta v_a$. Thus, torsion quantifies how the parallelogram is no longer closed by changing the order of parallel transport \citep{Bahamonde_2024}. Finally, the non-metricity tensor $Q_{\alpha\mu\nu}$ quantifies the change in a vector's length when comparing it at two infinitesimally close points $p$ and $q$ \citep{Heisenberg_2024}.

The most general connection includes the Levi-Civita standard connection, plus terms of the torsion and non-metricity tensors \citep{Bahamonde_2024, Heisenberg_2024}. We are free to put restrictions on the connection. Thus, we can set one or two of the three fundamental tensors to zero. There are in fact two equivalent ways of building GR in flat spacetime, which alongside GR leads to the \text{geometric trinity of gravity} \citep{Beltran_2019}. We have the Teleparallel Equivalent of General Relativity (TEGR), where we set the Riemann and non-metricity tensors to zero and gravity is driven by torsion \citep{Maluf_2013}. On the other hand, if we set the Riemann and torsion tensors to zero, we get the STEGR \citep{Nester_1998}. There is no observational difference between the three cases. However, their theoretical construction is quite different, as it completely changes how parallel transport works. By generalizing the GR Lagrangian, for instance by promoting $R \to f(R)$, we build a new set of theories where we have arbitrary functions of the Ricci scalar \citep{DeFelice_2010}. Similarly, we can build $f(T)$ \citep{Cai_2016} and $f(Q)$ \citep{Jimenez_2018} theories by having arbitrary functions of the torsion and non-metricity scalars, respectively. In the present paper, we focus on the latter formalism, called \textit{symmetric teleparallel gravity}. Symmetric means the connection is symmetric in its lower indices, while teleparallel means the curvature is zero.

The fundamental tensor in symmetric teleparallelism is the non-metricity tensor, which is given by the covariant derivative of the metric
\begin{equation}
    Q_{\alpha\mu\nu} \equiv \nabla_\alpha g_{\mu\nu},
\end{equation}
from which we can derive several tensors called the disformation and superpotential \citep{Jimenez_2018, Jimenez_2020}.
\begin{equation}
    L^\alpha_{\;\;\mu\nu} = \frac{1}{2} Q^\alpha_{\;\;\mu\nu} - Q^{\;\;\;\alpha}_{(\mu \;\;\;\nu)},
\end{equation}
\begin{equation}
    P^\alpha_{\;\;\mu\nu} = -\frac{1}{2} L^\alpha_{\;\;\mu\nu} + \frac{1}{4} (Q^\alpha - \tilde{Q}^\alpha) g_{\mu\nu} - \frac{1}{4} \delta^\alpha_{(\mu} Q_{\nu)},
\end{equation}
where $Q_\alpha = g^{\mu\nu} Q_{\alpha\mu\nu}$ and $\tilde{Q}_\alpha = g^{\mu\nu} Q_{\mu\alpha\nu}$. Finally, the non-metricity scalar is \citep{Jimenez_2018, Jimenez_2020}
\begin{equation}
    Q = - Q_{\alpha\mu\nu} Q^{\alpha\mu\nu} = -\frac{1}{4} (-Q^{\alpha\mu\nu} Q_{\alpha \mu \nu} + 2 Q^{\alpha \mu \nu} Q_{\nu \alpha \mu} - 2 Q^\alpha \tilde{Q}_\alpha + Q^\alpha Q_\alpha),
\end{equation}
so the non-metricity scalar is given by four traces of the non-metricity tensor.

We can now consider the general action 
\begin{equation}
    S = \int d^4x \sqrt{-g} \left( -\frac{1}{2} f(Q) + \mathcal{L}_m \right),
\end{equation}
where $g$ is the determinant of the metric, $f(Q)$ is a function of the non-metricity scalar, and $\mathcal{L}_m$ is the matter Lagrangian. STEGR takes the form $f(Q) = Q/(8\pi G)$ \citep{Jimenez_2018}. By taking the variation $\delta S$ with respect to the metric ($\delta S/\delta g^{\mu\nu}$) and setting it to zero, we derive the field equations \citep{Jimenez_2018}
\begin{equation}
    \frac{2}{\sqrt{-g}} \nabla_\alpha \left[ \sqrt{-g} P^\alpha_{\;\;\mu\nu} f_Q  \right] + f_Q q_{\mu\nu} - \frac{1}{2} f g_{\mu\nu} = T_{\mu\nu},
\end{equation}
where $f_Q = df/dQ$ and \begin{equation}
    q_{\mu\nu} = P_{(\mu|\alpha\beta} Q_{|\nu)}^{\;\;\,\alpha\beta} - 2P^{\alpha\beta}_{(\nu|} Q_{\alpha\beta|\mu)}.
\end{equation} 

Finally, by taking $\delta S/\delta \Gamma^{\alpha}_{\;\;\mu\nu} = 0$, we get \citep{Jimenez_2018}
\begin{equation}
    \nabla_\mu \nabla_\nu (\sqrt{-g} f_Q P^{\mu\nu}_{\;\;\;\,\alpha}) = 0.
\end{equation}

\subsection{Modified Friedmann equations}
\label{sec:ModifiedFriedmanneqs}

To derive the modified Friedmann equations, we start from the flat Friedmann-Lema{\^\i}tre-Robertson-Walker \citep[FLRW;][]{Friedmann_1922, Friedmann_1924, Lemaitre_1931, Robertson_1935, Robertson_1936a, Robertson_1936b, Walker_1937} metric
\begin{equation}
    ds^2 = - dt^2 + a^2(t) \delta_{ij} dx^i dx^j. 
\end{equation}
In this case, we are setting the lapse function $N(t) = 1$. From a general lapse function, we can always take the time reparametrization $t = \int_0^t d\tau \sqrt{-g_{tt}(\tau)}$, i.e., the lapse function is unphysical \citep{Ambrosio_2022}.

The remaining ingredient to derive the Friedmann equations is the connection. The most general connection for a flat spacetime in the symmetric teleparallel framework is given by \citep{Ambrosio_2022, Hohmann_2021}
\begin{equation}
   \Gamma^t_{\;\;\mu\nu} =  \begin{pmatrix}
        C_1 & 0 & 0 & 0 \\
        0 & C_2 & 0 & 0 \\
        0 & 0 & C_2 r^2 & 0  \\
        0 & 0 & 0 & C_2 r^2\sin \theta
    \end{pmatrix},
\end{equation}
\begin{equation}
    \Gamma^r_{\mu\nu} = \begin{pmatrix}
        0 & C_3 & 0 & 0 \\
        C_3 & 0 & 0 & 0 \\
        0 & 0 & -r & 0 \\
        0 & 0 & 0 & -r\sin^2\theta
    \end{pmatrix},
\end{equation}
\begin{equation}
    \Gamma^\theta_{\;\;\mu\nu} = \begin{pmatrix}
        0 & 0 & C_3 & 0 \\
        0 & 0 & \dfrac{1}{r} & 0 \\
        C_3 & \dfrac{1}{r} & 0 & 0 \\
        0 & 0 & 0 & -\cos\theta\sin\theta
    \end{pmatrix},
\end{equation}
\begin{equation}
    \Gamma^\phi_{\;\;\mu\nu} = \begin{pmatrix}
        0 & 0 & 0 & C_3 \\
        0 & 0 & 0 & \dfrac{1}{r} \\
        0 & 0 & 0 & \cot \theta \\
        C_3 & \dfrac{1}{r} & \cot\theta & 0
    \end{pmatrix},
\end{equation}
where $C_1(t)$, $C_2(t)$, and $C_3(t)$ are functions of time. There are three different possible forms for these functions that satisfy spatial flatness in the symmetric teleparallel framework. These are \citep{Ambrosio_2022}:
\begin{enumerate}
    \item Connection I: $C_1(t) = C_3(t) + \dfrac{\dot{C_3}(t)}{C_3(t)}, C_2(t) = 0$.
    \item Connection II: $C_1(t) = -\dfrac{\dot{C_2}(t)}{C_2(t)}, C_3(t) = 0$.
    \item Connection III: $C_2(t)=C_3(t)=0$. 
\end{enumerate}
Connection III is the simplest as the function $C_1(t)$ does not appear in the modified Friedmann equations and it is undetermined by them \citep{Ambrosio_2022}. Furthermore, by setting $C_1(t) = 0$, we recover the so-called \textit{coincident} gauge, where the connection vanishes and the covariant derivatives become partial derivatives \citep{Jimenez_2018}. Connections I and II allow for an additional degree of freedom that changes the dynamics of the Hubble factor. For example, different connections imply different cosmological constraints \citep{Shi_2023} and different potential stability concerns \citep{Beltran_2021, Aguiar_2024, Bello_2024, Narawade_2025}. Furthermore, different connections can resolve cosmological singularities like the Big Bang and Big Rip, replacing them with smooth de-Sitter phases \citep{Ayuso_2025}. Therefore, considering different connections gives more flexibility to $f(Q)$ gravity models, with the potential of alleviating or solving problems in cosmology like the Hubble and BAO tensions.

In this paper, we will work with Connection III as it is the one that has been studied the most. This will provide insights on the capabilities of $f(Q)$ gravity to solve the Hubble tension without the flexibility of an extra degree of freedom. Future work might consider alleviating the Hubble tension within $f(Q)$ with the assistance of a connection degree of freedom that can increase the accelerated expansion at late times and simultaneously fit the BAO data. 

By using Connection III and considering a perfect fluid with density $\rho$ and pressure $P$, the modified Friedmann equations are
\begin{equation}
    Qf_Q - \frac{1}{2} f = \rho,
\end{equation}
\begin{equation}
    (2Qf_{QQ} +f_Q) \dot{H} = -\frac{1}{2} (\rho + P),
\end{equation}
\begin{equation}
    Q = 6H^2,
\end{equation}
\begin{equation}
    \dot{\rho} + 3H(\rho + P) = 0.
\end{equation}
By using the STEGR form $f(Q)= Q/(8\pi G)$, we recover the standard Friedmann equations. If we add these on, we can separate the density and pressure contributions from the effective $f(Q)$ gravity contributions. This allows us to derive the effective $f(Q)$ equation of state (EoS)
\begin{equation}
\label{eqn:fQ_EoS}
    w_{f(Q)} = \left(\frac{\kappa p(1-\tilde{f}_Q-12H^2\tilde{f}_{QQ}) + 3H^2 \tilde{f}_Q - 36 H^4 \tilde{f}_{QQ} - \dfrac{1}{2}\tilde{f}}{\tilde{f}_Q + 12H^2 \tilde{f}_{QQ}} \right)\left(\frac{2 \tilde{f}_Q}{\dfrac{1}{2} \tilde{f} + \kappa \rho (1-2 \tilde{f}_Q)}\right),
\end{equation}
where $\kappa \equiv 8\pi G$, $f \equiv \tilde{f}/\kappa$, $f_Q \equiv \tilde{f}_Q/\kappa$, and $f_{QQ} \equiv \tilde{f}_{QQ}/\kappa$. This expression is useful to determine whether the effective $f(Q)$ dark energy has quintessence, cosmological constant, phantom, or phantom crossing behaviour. 

\section{Specific \texorpdfstring{$f(Q)$}{f(Q)} Models}
\label{sec:fQ_models}

In this section, we present the models we study. We explore three models previously studied in the literature that geometrically imply an accelerated expansion of the universe and include additional terms that can increase $H(z)$ at late times. The goal is to see if they can solve the Hubble tension and achieve consistency with the DESI~DR2 BAO data \citep{DESI_2025}. This is important as doing both is difficult for a cosmological model. For example, the Chevallier-Polarski-Linder \citep[CPL;][]{Chevallier_2001, Linder_2003, Linder_2008} $w_0 w_a$ parametrization (with $w$ depending linearly on $a$) is preferred over $\Lambda$CDM at the $3.1\sigma$ C.L. using DESI~DR2 and CMB data \citep{DESI_2025, Cortes_2025}. However, the Bayesian Evidence (Eq.~\ref{eq:Bayesian_Evidence}) disfavours the addition of two extra parameters to reduce the BAO tension \citep{Ong_2025}. Moreover, the CPL model predicts a low value for the Hubble constant of $H_0 = 63.9^{+1.7}_{-2.1}$ km/s/Mpc using CMB + DESI~DR2 \citep{Mirpoorian_2025}. This is in $4.86\sigma$ tension with the latest independent $H_0$ determination using the Cepheid-SNIa route by the SH0ES Collaboration \citep{Breuval_2024}. In physically motivated models, $w$ must have a specific redshift dependence to solve the Hubble tension, but this can be difficult to reconcile with the latest BAO data \citep{Lee_2022_DE}.

To clarify how the standard Friedmann equations are modified in our considered models, we define $\tilde{E}$ as the value of $E(z) \equiv H(z)/H_0$ in $\Lambda$CDM, with $\Lambda \neq 0$ or $\Lambda = 0$ subscripts used to indicate if the dark energy contribution is included.
\begin{equation}
    \tilde{E}^2_{\Lambda \neq 0} = \Omega_\Lambda + \Omega_{bc} (1+z)^3 + \Omega_\gamma (1+z)^4 + \Omega_{\nu, ur} (1+z)^4 \sqrt{1+\dfrac{1}{a^2_{nr}(1+z)^2}},
    \label{eq:Friedmann_with_Lambda}
\end{equation}
where $\Omega_x$ refers to the present fraction of the cosmic critical density $3H_0^2/(8\pi G)$ in component $x$ (which is $\Lambda$, $bc$, $\gamma$, or $\nu$ for dark energy, baryons + CDM, photons, or neutrinos, respectively), $\Omega_{\nu, ur}$ is the present neutrino contribution if the neutrinos were massless, and $a_{nr}$ is the scale factor when the neutrinos transition from ultra-relativistic to non-relativistic (Eq.~\ref{eqn:a_nr}). We only consider flat models, neglecting the curvature contribution. The subscript $\Lambda \neq 0$ indicates that the canonical dark energy contribution is included. We initially consider the simpler $\Lambda = 0$ case in which the $\Omega_\Lambda$ term on the right hand side is excluded. At the background level, models other than $\Lambda$CDM can be specified through their relation between $E \equiv H/H_0$ and $\widetilde{E}$ with or without $\Lambda$, depending on the model.

\subsection{Logarithmic \texorpdfstring{$f(Q)$}{f(Q)} Model (``Log'')}

First, we propose a logarithmic model of the form
\begin{equation}
    f(Q) = \frac{Q}{8\pi G} - \alpha \ln \left( \frac{Q}{Q_0} \right),
\end{equation}
with $Q_0 \equiv Q(z=0)$. This model predicts accelerated expansion consistently with SNIa and BAO data \citep{Najera_2023}. Those authors show that the Friedmann equation for $E$ is
\begin{equation}
\label{eqn:FriedmannLogarithmic}
    E^2 + \Omega_\alpha \ln E = \Omega_\alpha + \tilde{E}^2_{\Lambda = 0}(z), 
\end{equation}
where $\Omega_\alpha \equiv (8\pi G \alpha)/(3H_0^2)$ is an effective geometric density. This definition ensures that Eq.~\ref{eqn:FriedmannLogarithmic} satisfies the closure relation
\begin{equation}
    \Omega_\alpha + \tilde{E}^2_{\Lambda=0}(z=0) = 1,
\end{equation}
which is apparent by setting $z = 0$ and $E = 1$. The model has the same number of parameters as $\Lambda$CDM. As we can see, the Friedmann equation is the $\Lambda$CDM one apart from a logarithmic correction to the left-hand side. When $z \gg 1$, this term is much smaller than $E^2$, making the behaviour the same as $\Lambda$CDM. This is in line with the lack of reported deviations from $\Lambda$CDM at high $z$, with the DESI BAO data at $z > 2$ in especially good agreement with the \emph{Planck} cosmology \citep{Banik_2025_BAO}. However, at low $z$, the logarithmic term makes an additional contribution that can enhance the late-time accelerated expansion.

\subsection{Exponential \texorpdfstring{$f(Q)$}{f(Q)} Model (``Exp'')}

Next we consider an exponential model of the form
\begin{equation}
    f(Q) = \frac{Q}{8\pi G} \exp{\left( \frac{\lambda Q_0}{Q} \right)}.
\end{equation}
This was the first $f(Q)$ model that could challenge $\Lambda$CDM statistically \citep{Fotios_2021}. Its Friedmann equation is
\begin{equation}
\label{eqn:exponentialFriedmann}
    (E^2 - 2\lambda) \exp \left(\frac{\lambda}{E^2}\right) = \tilde{E}^2_{\Lambda=0}(z).
\end{equation}
The closure relation gives a solution for the $\lambda$ parameter:
\begin{equation}
    \lambda = \frac{1}{2} + \mathcal{W}_0 \left( - \frac{\tilde{E}^2_{\Lambda=0}(z=0)}{2 \sqrt{e}} \right),
\end{equation}
where $\mathcal{W}_0$ is the principal branch of the Lambert function. Therefore, this function also has the same number of parameters as $\Lambda$CDM. In this case, the modified Friedmann equation provides an exponential term that scales like $E^{-2}$. When $E^2 \gg \lambda$, then $E^2 \to \Omega_{bc} (1+z)^3 - \lambda$, having a behaviour like $\Lambda$CDM -- but with an exponential correction at low redshift.

\subsection{Hyperbolic Tangent \texorpdfstring{$f(Q)$}{f(Q)} Model (``Tanh'')}

Finally, we consider a third $f(Q)$ model given by \citep{Fotios_2023}:
\begin{equation}
    f(Q) = \frac{Q}{8\pi G} + \alpha \tanh \left( \frac{Q_0}{Q} \right),
\end{equation}
where the component enhancing the accelerated expansion comes from the hyperbolic tangent term. The Friedmann equation is 
\begin{equation}
    E^2 - \frac{\Omega_\alpha}{E^2} \left( \cosh^{-2}\left( E^{-2} \right) \right) - \frac{1}{2} \Omega_\alpha \tanh(E^{-2}) = \tilde{E}^2_{\Lambda = 0}(z),
\end{equation}
where $\Omega_\alpha = (8\pi G \alpha)/(3H_0^2)$, like in the logarithmic model. The closure relation gives
\begin{equation}
    \Omega_\alpha \left(\cosh^{-2}(1) + \dfrac{1}{2} \tanh(1) \right) + \tilde{E}^2_{\Lambda=0}(z=0) = 1,
\end{equation}
which again has the same number of parameters as $\Lambda$CDM.

\subsection{Including a Cosmological Constant}
\label{sec:including_Lambda}

In $f(Q)$ gravity, the late-time accelerated expansion of the Universe can arise purely from geometric effects, without the need for an exotic dark energy component \citep{Mandal_2020_energy, Visser_2000}. The non-linear dependence of the action on the non-metricity scalar $Q$ introduces effective gravitational terms that behave as a fluid with negative pressure. When these geometric contributions are moved to the right-hand side of the field equations, they form an effective energy--momentum tensor with $\rho_{\text{eff}} + 3P_{\text{eff}} < 0$, corresponding to an effective equation-of-state parameter $w_{\text{eff}} = P_{\text{eff}} / \rho_{\text{eff}} < -1/3$. This violation of the strong energy condition allows gravity to act repulsively and drive the observed cosmic acceleration. At the same time, viable $f(Q)$ models typically satisfy the weak and dominant energy conditions $\rho_{\text{eff}} > 0$, $\rho_{\text{eff}} + P_{\text{eff}} > 0$, and $\rho_{\text{eff}} \ge |P_{\text{eff}}|$, ensuring positive energy density, physically well-behaved matter, and causal (subluminal) energy flow, respectively. Thus, $f(Q)$ gravity provides a geometric explanation for cosmic acceleration while maintaining physically reasonable energy and causality properties \cite{Visser_2000}.

Despite the cosmological constant not being necessary to produce accelerated expansion in $f(Q)$ gravity, we also consider models in which it is included to gauge its effect on the Hubble and BAO tensions. This is achieved by adding $\Lambda/(4\pi G)$ to the $f(Q)$ function. The Friedmann equations get an additional $\Omega_\Lambda$ term on the right-hand side, while the closure relations get this term on the left-hand side. For the exponential model, the solution for $\lambda$ is now
\begin{equation}
    \lambda = \frac{1}{2} + \mathcal{W}_0 \left( - \frac{\tilde{E}_{\Lambda \neq 0}^2(z=0)}{2\sqrt{e}} \right).
\end{equation}
These variations have one more free parameter than $\Lambda$CDM. The new free parameter allows for extra flexibility in the Hubble parameter that might enhance the agreement with BAO measurements. This effectively restores the $\Lambda$CDM model plus an $f(Q)$ correction term, which gets a fine-tuning role. The drawback is that these models will only explain part of the dark energy geometrically, undermining their theoretical motivation.
We refer to these variants in the remainder of the paper as Log + $\Lambda$, Exp + $\Lambda$, and Tanh + $\Lambda$.

\subsection{Phenomenological Models (Phen)}

In addition to the $f(Q)$ models, we propose three flexible phenomenological models to study whether it is possible to solve the Hubble constant tension and BAO anomalies at the background level, and if so, what form such a solution would take. The proposed models are
\begin{equation}
\label{eqn>H_z_phen}
    H(z) = H_{0 \Lambda CDM} \tilde{E}_{\Lambda \neq 0}(z) + \Delta H(z; A, z_1),
\end{equation}
where $A$ and $z_1$ are model parameters. We consider these three $\Delta H(z)$ functions:
\begin{equation}
    \Delta H(z) = \begin{cases}
        A\,\exp \left( -\dfrac{z}{z_1} \right), \\
        \dfrac{A}{\cosh \left( \dfrac{z}{z_1} \right)}, \\
        \dfrac{A}{1 + \dfrac{z}{z_1}}.
        \end{cases}
    \label{eq:Delta_H_phen}
\end{equation}
We refer to these models as ``Phen, exp'', ``Phen, sech'' and ``Phen, PD'' (where PD stands for polynomial decay).

These models have four free parameters $(\Omega_{bc}, H_0, A, z_1)$, with the extra $\Delta H$ term introducing a correction that allows for faster accelerated expansion at late times. The predicted Hubble constant is $H_0 = H_{0 \Lambda CDM} + A$. The aim is to match the local $H_0$ measurements while restricting deviations from $\Lambda$CDM behaviour to $z \lesssim z_1$, which is found empirically but limited to $z_1 < 1$ as we are aiming for a late Universe solution.

\section{Observational Data}
\label{sec:ObsData}

We constrain the parameter vector $\Theta = \{ \Omega_m, H_0 \}$ for $\Lambda$CDM and the $f(Q)$ models, $\Theta = \{ \Omega_m, \Omega_\Lambda, H_0 \}$ for the cosmological constant variants, and $\Theta = \{ \Omega_m, H_0, A, z_1 \}$ for the phenomenological models, where $\Omega_m = \Omega_{bc} + \Omega_\nu$ is intended for comparison with late-time probes like BAO, when the neutrinos behave like matter. For the CMB likelihood, we add the parameter $w_b \equiv \Omega_b h^2$ as it is constrained by the CMB alongside $w_c \equiv \Omega_{c}h^2$. Since $w_b$ is tightly constrained by the CMB prior, we do not report its posteriors, but instead marginalise over it. To determine if the $f(Q)$ models can alleviate or solve the Hubble and BAO tensions, we use four of the latest independent $H_0$ constraints, the latest DESI~DR2 BAO \citep{DESI_2025}, and the $H(z)$ measurements obtained with the CC method \citep[][and references therein]{Moresco_2022}. Finally, we use the standard CMB constraints for the physical baryon density parameter $w_b$, the physical baryon + CDM density parameter $w_{bc}$, and the recombination acoustic angular scale $\theta_\star$ (see Appendix~\ref{sec:CMB-detailed}). If we want a solution to the Hubble tension, we need to either fit the CMB constraints and solve the tension at late times, or modify early time physics. The latter implies the age of the universe $\propto H_0^{-1}$, reducing its predicted age and causing additional complications fitting the CMB data \citep{Vagnozzi_2021, Vagnozzi_2023, Calabrese_2025, Camphuis_2025}. In the present paper, we choose the former approach and thus impose the standard CMB constraints on the cosmological parameters. We summarize the observational data in Table~\ref{tab:obsData}. We note that if the baryon and CDM densities at any $z$ and the comoving distance to recombination are the same as in the \emph{Planck} $\Lambda$CDM cosmology, there should be no difficulty fitting the CMB anisotropies given the lack of new physics at early times.

\begin{table}
\caption{Summary of the observational datasets considered in this paper.}
\begin{tabular}{|c|c|c|}
\hline
Data & Description & References \\
\hline
Local $H_0$ & Four model-independent local $H_0$ determinations & \citep{Breuval_2024, Pesce_2020, Jensen_2025, Vogl_2025} \\
\hline
DESI~DR2 BAO & The most recent BAO measurements from the DESI collaboration & \citep{DESI_2025} \\
\hline
Cosmic Chronometers & 32 $H(z)$ measurements using the differential age method in passive galaxies & \citep{Moresco_2020} \\
\hline
CMB constraints & \emph{Planck} 2018 + SPT-3G + ACT DR6 constraints for $(100\, \theta_\star, w_b, w_{bc})$ & \citep{Planck_2020, Camphuis_2025, Louis_2025} \\
\hline
\end{tabular}
\label{tab:obsData}
\end{table}

\subsection{Model-Independent \texorpdfstring{$H_0$}{H\_0} Determinations}
\label{sec:H0_estimates}

We use four recent model-independent $H_0$ constraints: the latest SH0ES constraint using the Small Magellanic Cloud as an extra anchor \citep{Breuval_2024}, the result from the Megamaser Cosmology project using only geometric distances \citep{Pesce_2020}, the result from the TRGB-SBF project using neither Cepheids nor SNIa \citep{Jensen_2025}, and a one-step (or rung-free) determination using Type~II supernovae \citep{Vogl_2025}. It can be argued that the model-independent $H_0$ determinations are in tension with the CMB data. However, there is no \emph{intrinsic} tension between the CMB and local $H_0$: the tension is model-dependent. A model which can solve the Hubble tension must simultaneously fit the CMB and the local $H_0$. Therefore, to solve the Hubble tension, we have to fit our models to both CMB+$H_0$, in addition to considering other datasets like BAO and CC.

\subsubsection{SH0ES}

The Supernova $H_0$ for the Equation of State (SH0ES) collaboration provides the most precise model-independent determination of the Hubble constant. This uses a three rung cosmic distance ladder, with geometric distances calibrating the Period-Luminosity relation of Cepheid variable stars \citep[the Leavitt Law;][]{Leavitt_1912}. The second rung uses this relation to determine the absolute magnitude of SNIa. Finally, the third rung uses SNIa in the Hubble flow with $z = 0.023 - 0.15$ to obtain that $H_0 = 73.17 \, \pm \, 0.86$ km/s/Mpc \citep{Breuval_2024}. This is in $5.8\sigma$ tension with the CMB + flat $\Lambda$CDM constraint from \emph{Planck}~2018 \citep[$H_0 = 67.36 \, \pm \, 0.54$ km/s/Mpc;][]{Planck_2020}. The discrepancy can be due to unaccounted-for systematics in the SH0ES Cepheid-SNIa distance ladder, though the good agreement with a number of other techniques argues against this interpretation \citep[][and references therein]{H0DN_2025, Valentino_2025}. The other possibility is new physics in the form of an extension to $\Lambda$CDM.

The phenomenological models have a caveat that needs to be handled for this $H_0$ constraint. SH0ES computes $H_0$ by fitting $H(z)$ with a Taylor series (the cosmographic expansion). If $z_1 \lesssim  0.15$ and thus lies within the SH0ES redshift range, the predicted $H(z)$ for the phenomenological models will strongly curve in this range. The phenomenological model prediction for $H_0$ will then not correspond to what we would infer from the SH0ES distance ladder. Thus, for the phenomenological models, we compute the predicted $H(z)$ for all SNIa redshifts inside the third rung of the distance ladder. We fit the second-order cosmographic expansion using $q_0 = -0.55$, the value considered by SH0ES \citep{Riess_2022_comprehensive, Breuval_2024}. The extrapolated $H(z=0)$ of our cosmographic expansion is then compared with SH0ES. This makes our analysis more realistic by predicting the $H_0$ we would derive in the model from the SH0ES distance ladder approach. For simplicity, we neglect the third order jerk parameter due to the low redshifts involved. We note that our other models lack sharp features in the predicted $H(z)$ at low $z$, so this complication does not arise.

\subsubsection{The Megamaser Cosmology Project}

This is an independent determination of $H_0$ using geometric distances to six megamaser host galaxies (NGC~4258, UGC~3789, CGCG~074-064, NGC~6323, NGC~5765b and NGC~6264). The final result is $H_0 = 73.9 \, \pm \, 3.0$ km/s/Mpc \citep{Pesce_2020}, which exceeds \emph{Planck}~2018 by $2.15\sigma$ and SH0ES by only $0.23\sigma$. 

\subsubsection{The TRGB-SBF Project}

The third $H_0$ measurement we consider is given by the TRGB-SBF project \citep{Jensen_2025}. This is a three rung distance ladder like SH0ES. However, the second rung calibrator is the Tip of the Red Giant Branch (TRGB) instead of Cepheids. Finally, the third rung uses Surface Brightness Fluctuations (SBF) instead of SNIa. This provides an independent determination of $H_0 = 73.8 \pm \, 2.4$ km/s/Mpc \citep{Jensen_2025}, which exceeds \emph{Planck}~2018 by $2.62\sigma$ and SH0ES by only $0.25\sigma$.

\subsubsection{Type II SN}

The fourth and final independent $H_0$ constraint we consider is a new determination using no rungs. This comes from spectral modelling of Type~II SN using the tailored expanding photosphere method. The $H_0$ constraint is $74.9 \, \pm \, 1.9$ km/s/Mpc \citep{Vogl_2025}. Those authors mention that the systematic uncertainty $\lesssim$ the statistical uncertainty. To be conservative, we consider the systematic uncertainty to equal the statistical one, giving $H_0 = 74.9 \, \pm \, 2.7$ km/s/Mpc. This determination is $2.74\sigma$ above \emph{Planck}~2018 and only $0.61\sigma$ above SH0ES.

\subsubsection{\texorpdfstring{$H_0$}{H\_0} likelihood}

We model each of the above $H_0$ constraints as Gaussian likelihood terms:
\begin{equation}
    \ln \mathcal{L}_\text{$H_0$} = - \frac{1}{2} \, \sum_{i=1}^4 \ln\left(2\pi \sigma^2_{H_{0,i}}\right) - \frac{1}{2}\sum_{i=1}^4 \left(\frac{H_0 - H_{0,i}}{\sigma_{H_{0,i}}}\right)^2,
\end{equation}
where $H_{0,i}$ is the $i$-th determination of $H_0$, whose $1\sigma$ C.L. uncertainty is $\sigma_{H_{0,i}}$. We include both the $\chi^2$ term and the Gaussian normalisation for a computation of the Bayesian Evidence. The index $i$ runs over the four independent $H_0$ measurements considered previously. These measurements have been chosen to be as independent as possible, which we are assuming by summing the log-likelihoods. The Type II SN constraint is fully independent of the others. There is a small covariance between the Megamaser determination with SH0ES and the TRGB-SBF project because the latter two use the NGC~4258 megamaser as a calibrator for Cepheids and the TRGB, but we neglect this covariance for simplicity. SH0ES and TRGB-SBF are fully independent since they use different calibrators (Cepheids vs TRGB) and distance indicators in the Hubble flow (SNIa vs SBF).

\subsection{Cosmic Chronometers (CC)}


We consider the CC compilation presented in \citep{Moresco_2022}, which includes 32 $H(z)$ measurements. While this work was under review, a new and quite precise CC measurement appeared at $z = 0.12$ \citep{Wang_2026}, but we do not include it here.

By considering an FLRW universe, we can write the Hubble factor as
\begin{equation}
    H(z) = - \frac{\dot{z}}{1+z}.
\end{equation}
$\dot{z}$ can be measured with the differential age evolution of early-type galaxies \citep{Jimenez_2002}. This provides model-independent measurements of the Hubble factor. The logarithmic CC likelihood is
\begin{equation}
    \ln \mathcal{L}_{CC} = - \frac{1}{2} \sum_{i=1}^{32} \ln(2\pi \sigma^2_{CC, \, i}) - \frac{1}{2} \sum_{i=1}^{32} \left( \frac{H(z_i, \Theta) - H_{CC}(z_i)}{\sigma_{CC, \, i}} \right)^2,
\end{equation}
where $H(z_i, \Theta)$ is the predicted Hubble factor and $H_{CC} (z_i)$ is the observed Hubble factor from CC, with $1\sigma$ C.L. uncertainty of $\sigma_{CC, \, i}$. The covariance is small for this dataset, so we neglect it.

\subsection{CMB Constraints}

The models we consider have standard physics prior to recombination, since we focus on late-time solutions to the cosmic tensions. This means we can continue to apply the standard constraints on the physical baryon and CDM densities and the angular scale of the sound horizon at recombination. We therefore take the standard CMB constraints on these three derived parameters and their covariance matrix. For this, we work with the chains from \cite{Camphuis_2025}, which combines Planck 2018 \cite{Planck_2020}, SPT-3G \cite{Camphuis_2025}, and ACT DR6 \cite{Louis_2025}. We follow a similar approach to that proposed in \cite{Lemos_2023} and taken by DESI~DR2 \cite{DESI_2025}. This approach is known as the compressed CMB likelihood, which entails a focus on the parameter vector $(100\,\theta_\star, w_b, w_{bc})$, where
\begin{equation}
    w_b \equiv \Omega_b h^2, \quad w_{bc} \equiv w_b + w_c \equiv \overbrace{\left( \Omega_b + \Omega_c \right)}^{\Omega_{bc}} h^2,
\end{equation}
with $w_b$ ($w_c$) proportional to the physical baryon (CDM) density today. The acoustic angular scale at recombination is
\begin{eqnarray}
    \theta_\star &=& \frac{r_\star}{D_M(z_\star)}, ~\text{where} \\
    D_M(z_\star) &=& c \int_0^{z_\star} \frac{dz}{H(z)},
\label{eqn:comovingDistRec}
\end{eqnarray}
$z_\star$ is the recombination redshift (Eq.~\ref{eqn:z_rec}), $r_\star$ is the comoving sound horizon then, and $D_M(z_\star)$ is the comoving distance to recombination. The compressed CMB likelihood comes from marginalizing over the integrated Sachs-Wolfe effect, lensing distortions, and other late-time effects \cite{Lemos_2023}. This approach allows for a model-independent determination of the cosmological parameters provided there is no new physics at early times.

The comoving sound horizon at recombination ($r_\star$) is not modified in our approach, allowing us to use the method in \cite{DESI_2025}, which in turn uses the fitting formula in \cite{Hu_1996}. We provide a detailed explanation of the CMB likelihood in Appendix~\ref{sec:CMB-detailed}, especially the calculation of $z_\star$ and thus $r_\star$.

Once we have determined $w_b$, $w_{bc}$, and $\theta_\star$, we find the CMB contribution to the overall model log-likelihood, which is
\begin{equation}
    \ln \mathcal{L}_{CMB} = -\frac{1}{2} \ln(2\pi |\mathbf{C}_{CMB}|) - \frac{1}{2} \Delta \mu(100 \, \theta_\star, w_b, w_{bc})^{T} \mathbf{C}_{CMB}^{-1} \Delta \mu (100 \, \theta_\star, w_b, w_{bc}),
\label{eqn:CMB-likelihood}
\end{equation}
where $|\mathbf{C}_{CMB}|$ is the determinant of the $3 \times 3$ compressed CMB covariance matrix and $\Delta \mu \equiv \mu_\text{sim} (100 \, \theta_\star, w_b, w_{bc}) - \mu_\text{obs}(100 \, \theta_\star, w_b, w_{bc})$ is the difference between the predicted and observed values.

\subsection{DESI~DR2 BAO}

The BAO is a \emph{statistical} standard ruler arising from baryonic matter fluctuations in the primordial Universe \citep{Eisenstein_1998, Chen_2024}. We include the measurements from the latest release by the Dark Energy Spectroscopic Instrument \citep[DESI;][]{DESI_2025}. Their DR2 provides six measurements of $D_H/r_d$, six of $D_M/r_d$ at the same redshifts, and one of $D_V/r_d$, where
\begin{eqnarray}
    D_H(z) &=& \frac{c}{H(z)}, \\
    D_M(z) &=& c \int_0^z \frac{d\tilde{z}}{H(\tilde{z})},\\
    D_V(z) &\equiv& [z D^2_M(z) D_H(z)]^{1/3}.
\end{eqnarray}

The comoving sound horizon is
\begin{equation}
    r_d = \int_{z_d}^\infty \frac{c_s(\tilde{z})}{H(\tilde{z})} d\tilde{z},
\end{equation}
where $c_s$ is the sound speed. We are interested in studying late-time solutions to the Hubble tension, without introducing new physics at early times. Thus, we take the Brieden et al. formula for the sound horizon at the drag epoch \citep{Brieden_2023}:
\begin{equation}
    r_d = 147.05 \, \text{Mpc} \left( \frac{w_{bc}}{0.1432} \right)^{-0.23} \left( \frac{N_\text{eff}}{3.04} \right)^{-0.1} \left( \frac{w_b}{0.02236} \right)^{-0.13},
    \label{eqn:rd}
\end{equation}
where $N_\text{eff}$ is the effective number of neutrino species. The BAO contribution to the natural logarithmic likelihood is given by
\begin{equation}
    \ln \mathcal{L}_{\text{DESI~DR2}} = -\frac{1}{2} \ln(2\pi |\mathbf{C}_{\text{DESI~DR2}}|) - \frac{1}{2} \Delta G(z, \Theta)^T \mathbf{C}^{-1}_{\text{DESI~DR2}} \Delta G(z, \Theta),
\end{equation}
where $\mathbf{C}_{\text{DESI~DR2}}$ is the $13 \times 13$ covariance matrix of the data and $\Delta G(z, \Theta) \equiv G_\text{sim}(z, \Theta) - G_\text{DESI~DR2}(z)$ is the difference between predicted and observed values of the 13 DESI~DR2 measurements. Note that the covariance matrix is in block-diagonal form because observations at different redshifts use distinct galaxies and are therefore independent.

We only quantitatively compare our models to BAO measurements from DESI~DR2, but there are plenty of older BAO measurements over the last twenty years (table~1 in \citep{Banik_2025_BAO}). To give a broader overview of how well each model works, we visually compare our best-fitting models to this compilation of BAO data, mainly in Appendix~\ref{sec:FiguresBAO}.

\section{Results}
\label{sec:Results}

We use the nested sampling algorithm \texttt{dynesty}\footnote{\url{https://dynesty.readthedocs.io/en/v2.1.5/}} \citep{Speagle_2020, Koposov_2024, Skilling_2004, Skilling_2006, Feroz_2009} to compute the posterior probability and the Bayesian Evidence $Z$, given by
\begin{equation}
    Z(\mathcal{M}) = \int P(\theta) P(\mathbf{D}|\theta, \mathcal{M}) d\theta,
    \label{eq:Bayesian_Evidence}
\end{equation}
where $\theta$ is the parameter vector, $P(\theta)$ the prior, $P(\mathbf{D}|\theta, \mathcal{M})$ the likelihood, and $\mathcal{M}$ the model. Since the Evidence integrates over the full parameter space, models with more parameters get penalised as the larger parameter space reduces the fraction of the parameter space that is consistent with observations. This is a more rigorous implementation of the ancient idea that more complicated models should be penalised. We can approximately implement this simply by adding an ``Occam penalty'' directly to the likelihood of the best-fitting model \citep[e.g. see eq.~6 of][]{Haslbauer_2024}. However, the Bayesian Evidence provides a much more rigorous method to implement the idea of Occam's Razor.

We take 500 live points and use a stopping criterion of $\Delta \ln \hat{Z} = 0.01$, where the hat denotes that this is the approximate remaining Evidence, not the true Evidence. This tolerance is sufficient for our purposes because the different models considered have much larger Evidence differences. We plot the $1\sigma$ and 2$\sigma$ C.L. contours using the \texttt{GetDist} package \footnote{\url{https://getdist.readthedocs.io/en/latest/}} \citep{Lewis_2025_GetDist}. Posterior convergence was confirmed by checking that the number of effective samples was $N_{\mathrm{eff}} > 2000$ for all models.

To determine if the theoretical models can challenge the standard $\Lambda$CDM model, we compute their logarithmic Bayes factor. This is given by
\begin{equation}
    \ln B_{PB} = \ln Z_P - \ln Z_B,
\end{equation}
where $P$ is the proposed model and $B$ is the baseline model ($\Lambda$CDM). The Bayes factor will be negative if the $\Lambda$CDM model is preferred over the modified $f(Q)$ models, whereas the modified models are statistically preferred if the Bayes factor is positive. The strength of the preference is given by the Jeffreys criterion, as described by Kass and Raftery \citep{Kass_1995} following the convention often adopted in the cosmology literature \citep{Trotta_2008}. According to this, $|\ln B_{PB}| <1$=``weak evidence'', $1<|\ln B_{PB}| <3$=``substantial evidence'', $3<|\ln B_{PB}| <5$=``strong evidence'', and $|\ln B_{PB}| > 5$=``very strong evidence'' for the model with higher $Z$. However, readers are welcome to interpret our Bayesian Evidence differences using a different scale.


To quantify the significance of the BAO anomaly, we use two methods. For the first one, we make predictions for the DESI~DR2 data with model parameters constrained using the CMB alone, following a common convention \citep{DESI_2025, Camphuis_2025}. We run the MCMC with the CMB likelihood alone. Then, we take each step in the MCMC chain and predict the 13 DESI measurements. For each one, we get an array. We then take the weighted means and weighted covariance matrix, which we call $G_\text{DESI, predictions}$ and $\mathbf{C}_\text{DESI, predictions}$, respectively. We define $\Delta G_\text{BAO, DS}(z)= G_\text{DESI~DR2}(z) - G_\text{DESI, predictions}(z)$, where DS stands for \textit{data space}. We compute the $\chi^2$ as
\begin{equation}
    \chi^2_{\mathrm{DS}} = \Delta G_\text{BAO, DS}^T (\mathbf{C}_\text{DESI~DR2} + \mathbf{C}_\text{DESI, predictions})^{-1} \Delta G_\text{BAO, DS}.
\end{equation}
We convert this $\chi^2$ to a p-value for 13 degrees of freedom (d.o.f.) and derive the Gaussian equivalent tension with the inverse survival function. This is the data space BAO tension. We stress that this is not symmetric: predicting the DESI DR2 data using the CMB MCMC chain is very different to predicting the CMB on the basis of BAO data. Since the CMB has much more information than the BAO, predicting the CMB using the BAO is highly inaccurate. That is why we only predict the BAO data using the CMB and not vice versa.

For our second method, we run the MCMC separately with the CMB or BAO likelihood alone. We take the 2D parameter space $\Theta = (\Omega_m, H_0 \, r_d)$. We compute the weighted mean values and covariance for both datasets. The $\chi^2$ is defined as
\begin{equation}
    \chi^2_{\mathrm{PS}} = \Delta \Theta^T \left[ \mathbf{C}_\text{CMB}(\Theta) + \mathbf{C}_\text{DESI~DR2}(\Theta) \right]^{-1} \Delta \Theta,
\end{equation}
which is the same procedure used in \cite{DESI_2025, Camphuis_2025}. This quantifies a parameter space BAO tension for 2 d.o.f. Although this is less direct than the data space tension, it is much better suited to detecting systematic deviations from the model predictions, provided these deviations can be easily fit using revised $\Omega_m$ and/or $H_0 \, r_d$ (Appendix~\ref{BAO_toy_model}). We streamline our discussion of the BAO tension faced by any particular model by using the maximum of the data and parameter space (MDPS) tension, though we list both tensions for completeness. The MDPS tension is useful because it highlights if there is tension either directly in the data space or in the commonly used $(\Omega_m, H_0 \, r_d)$ parameter space. It is possible for a genuine anomaly to appear only in data space, if the anomaly cannot be captured by altering the parameters. But if it can, then the parameter space tension could be more useful, as we illustrate in Appendix~\ref{BAO_toy_model} using a simple toy model of the BAO anomaly.

It is important to highlight that in our computations of the BAO tension, we are taking a Gaussian approximation for the predictions and the $(\Omega_m, H_0 \, r_d)$ parameters. This assumption is accurate for $\Lambda$CDM, the $f(Q)$, and the $f(Q)+\Lambda$ models due to their limited flexibility, especially when model parameters are constrained using all non-BAO datasets rather than just the CMB. However, for the more flexible phenomenological models where the model parameters are more difficult to constrain, non-Gaussianity might slightly reduce the accuracy of the tension metrics. Even so, we expect the MDPS tension to give a good indication of whether a model could have predicted the BAO data on the basis of other datasets, mimicking the scientific method of checking if \emph{a priori} predictions are confirmed by subsequent observations.

\begin{figure}
\centering
\includegraphics[width=5.5cm]{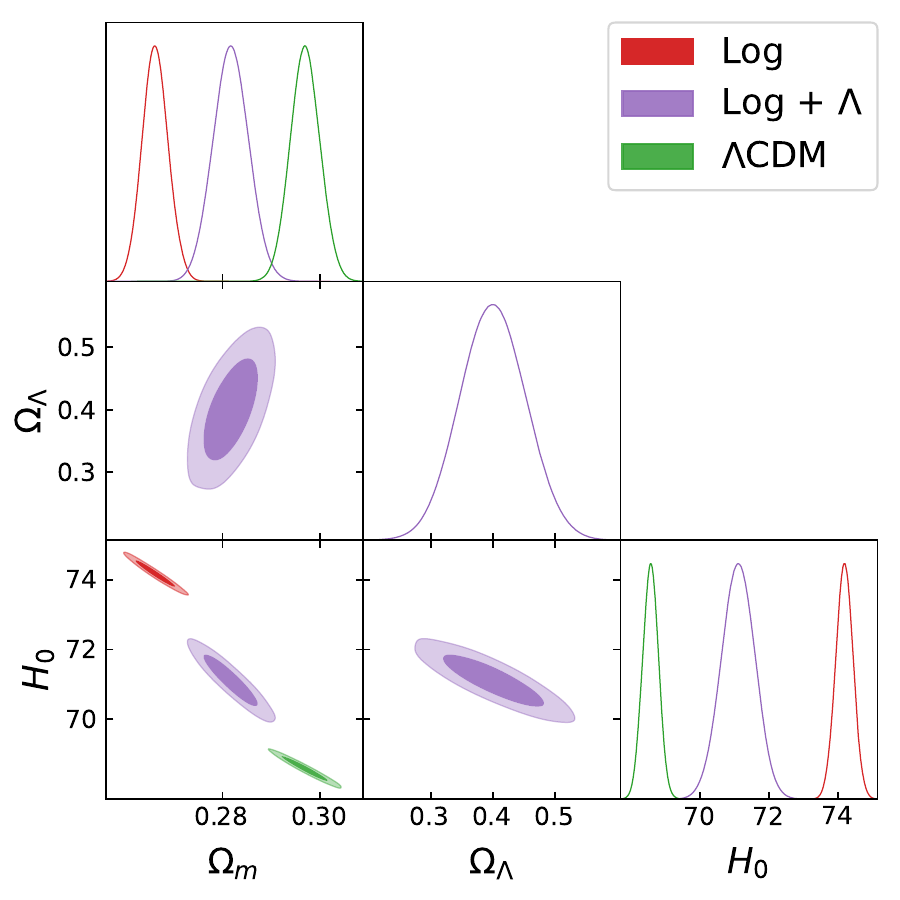}
\includegraphics[width=5.5cm]{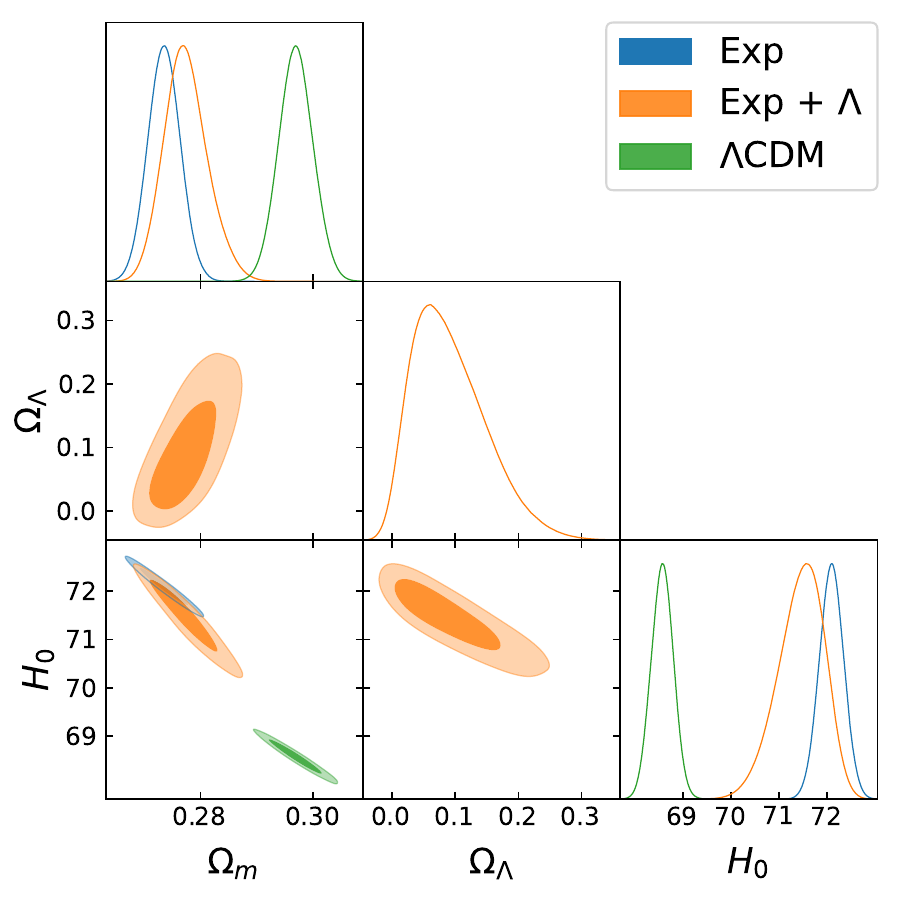}
\includegraphics[width=5.5cm]{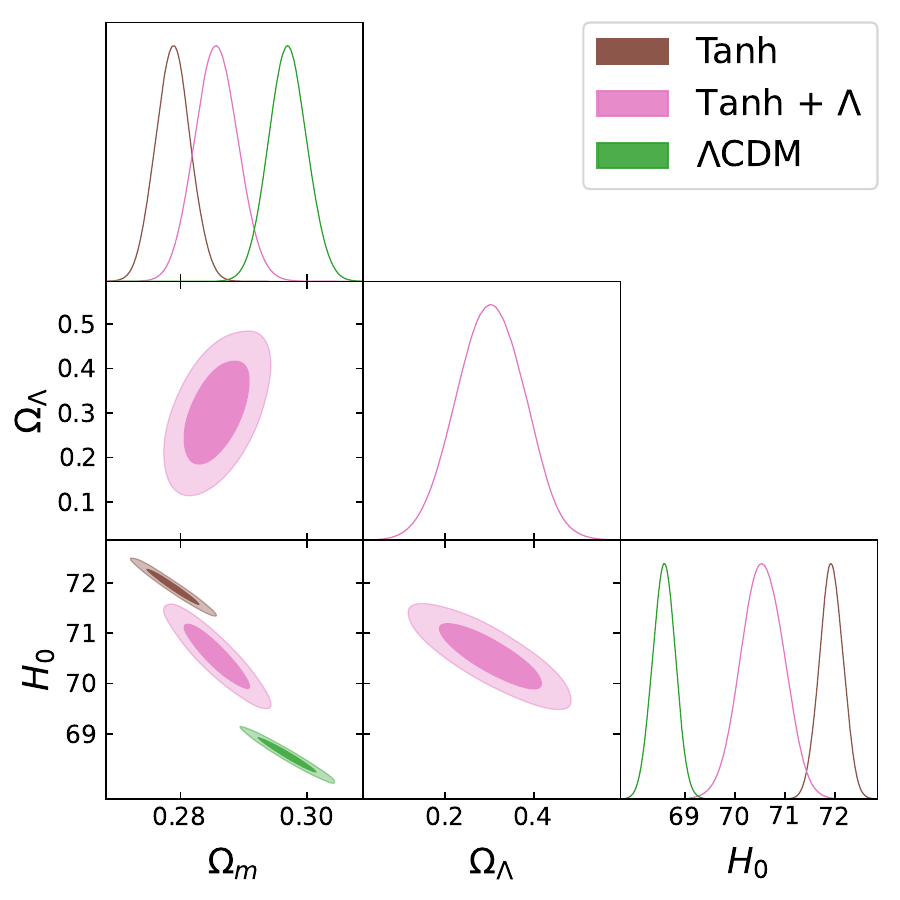}
\caption{$1\sigma$ and $2\sigma$ C.L. contours for the cosmological parameters $\Theta = \{ \Omega_m, \Omega_\Lambda, H_0$\}. Each panel shows results for a particular $f(Q)$ gravity model, the same model assisted by the cosmological constant $\Lambda$, and $\Lambda$CDM. We show (\textbf{a}) the logarithmic form $f(Q) = Q/(8 \pi G) - \alpha \ln(Q/Q_0)$, (\textbf{b}) the exponential form $f(Q) = Q/(8\pi G) \times \exp(\lambda Q_0/Q)$, and (\textbf{c}) the hyperbolic tangent form $f(Q) = Q/(8\pi G) + \alpha \tanh (Q_0/Q)$. Notice that once $\Lambda$ is allowed, only the Exp model can plausibly manage without it.}
\label{fig:f_Q_contours}
\end{figure}

\begin{figure}
\centering
\includegraphics[width=9.0cm]{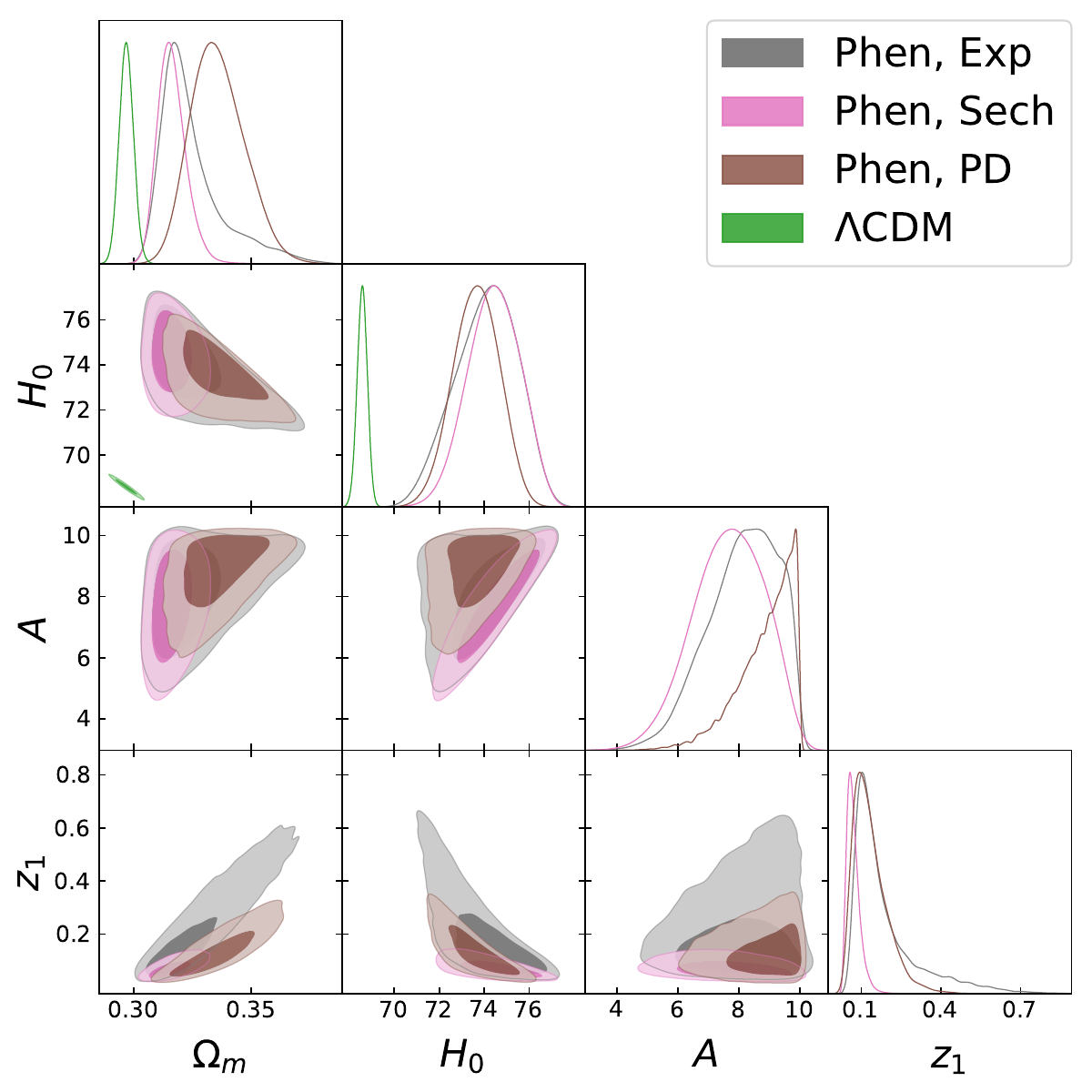}
    \caption{$1\sigma$ and $2\sigma$ C.L. contours for the cosmological parameters $\Theta = \{\Omega_m, H_0, A, z_1\}$ in the phenomenological models where $H_{\Lambda CDM}(z)$ is supplemented by an extra term at $z \lesssim z_1$ (Eq.~\ref{eq:Delta_H_phen}). We also show the $\Lambda$CDM model for comparison. Since this underestimates $H_0$ by $\approx 6$~km/s/Mpc, we limit $A$ to the range $(0, 10)$~km/s/Mpc.
    }
    \label{fig:phen_contours}
\end{figure}

As we can see in Table~\ref{tab:ResultsfQ} and Figure~\ref{fig:f_Q_contours}, all the modified $f(Q)$ models are capable of solving -- or at least significantly ameliorating -- the Hubble tension by giving high values for $H_0$ of around 72 km/s/Mpc, or even 74 km/s/Mpc for the Log model. However, when the cosmological constant is added, the fit to the local $H_0$ measurements worsens, going below 72 km/s/Mpc in all cases. Since this preference cannot be driven by the high local $H_0$ measurements (section~\ref{sec:H0_estimates}), it most likely comes from the BAO, an issue we will return to. The phenomenological models also resolve the Hubble tension, having $H_0 \approx 74$~km/s/Mpc. Interestingly, $\Omega_m$ is higher for these models (Table~\ref{tab:ResultsPhen} and Figure~\ref{fig:phen_contours}).

\begin{table}
    \caption{Mean values and $1\sigma$ C.L. uncertainties for the parameter vector $\Theta = \{ \Omega_m, \Omega_\Lambda, H_0 \}$ in the $\Lambda$CDM and $f(Q)$ models. Notice that the Exp model does not need $\Lambda$. 
    }
    \begin{tabular}{|c|c|c|c|}
    \hline
    \textbf{Model}	& $\Omega_m$ & $\Omega_\Lambda$ & $H_0$ (km/s/Mpc) \\
    \hline
    $\Lambda$CDM & $0.2969\pm 0.0030$ & $1 - \Omega_m - \Omega_\gamma$ & $68.58\pm 0.23$ \\ [5pt]
    Log & $0.2662\pm 0.0027$ & - & $74.18\pm 0.25$ \\
    Log + $\Lambda$ & $0.2817\pm 0.0036$ & $0.400\pm 0.053$ & $71.11\pm 0.48$ \\ [5pt]
    Exp & $0.2736\pm 0.0028$ & - & $72.09\pm 0.25$ \\
    Exp + $\Lambda$ & $0.2774^{+0.0034}_{-0.0041}$ & $0.091^{+0.040}_{-0.071}$ & $71.46^{+0.54}_{-0.41}$ \\ [5pt]
    Tanh & $0.2788\pm 0.0027$ & - & $71.92\pm 0.24$ \\
    Tanh + $\Lambda$ & $0.2857\pm 0.0034$ & $0.302\pm 0.076$ & $70.55\pm 0.42$ \\
    \hline
    \end{tabular}
    \label{tab:ResultsfQ}
\end{table}

\begin{table}
    \caption{Mean values and $1\sigma$ C.L. uncertainties for the parameter vector $\Theta = \{ \Omega_m, H_0, A, z_1 \}$ in the phenomenological models.}
    \begin{tabular}{|c|c|c|c|c|c|c|}
    \hline
    \textbf{Model}	& $\Omega_m$ & $H_0$ (km/s/Mpc) & $A$ (km/s/Mpc) & $z_1$ \\
    \hline
    $\Lambda$CDM & $0.2969\pm 0.0030$ & $68.58\pm 0.23$ & - & - \\
    Phen, exp & $0.3246^{+0.0049}_{-0.015}$ & $74.1^{+1.6}_{-1.2}$ & $8.13^{+1.6}_{-0.72}$ & $0.190^{+0.016}_{-0.13}$ \\
    Phen, sech & $0.3164^{+0.0047}_{-0.0068}$ & $74.4\pm 1.2$ & $7.7^{+1.3}_{-1.1}$ & $0.071^{+0.013}_{-0.030}$ \\
    Phen, PD & $0.3358^{+0.0098}_{-0.013}$ & $73.6\pm 1.0$ & $8.86^{+1.1}_{-0.32}$ & $0.134^{+0.033}_{-0.079}$ \\
    \hline
    \end{tabular}
    \label{tab:ResultsPhen}
\end{table}

\begin{figure}
\centering
\includegraphics[width=5.5cm]{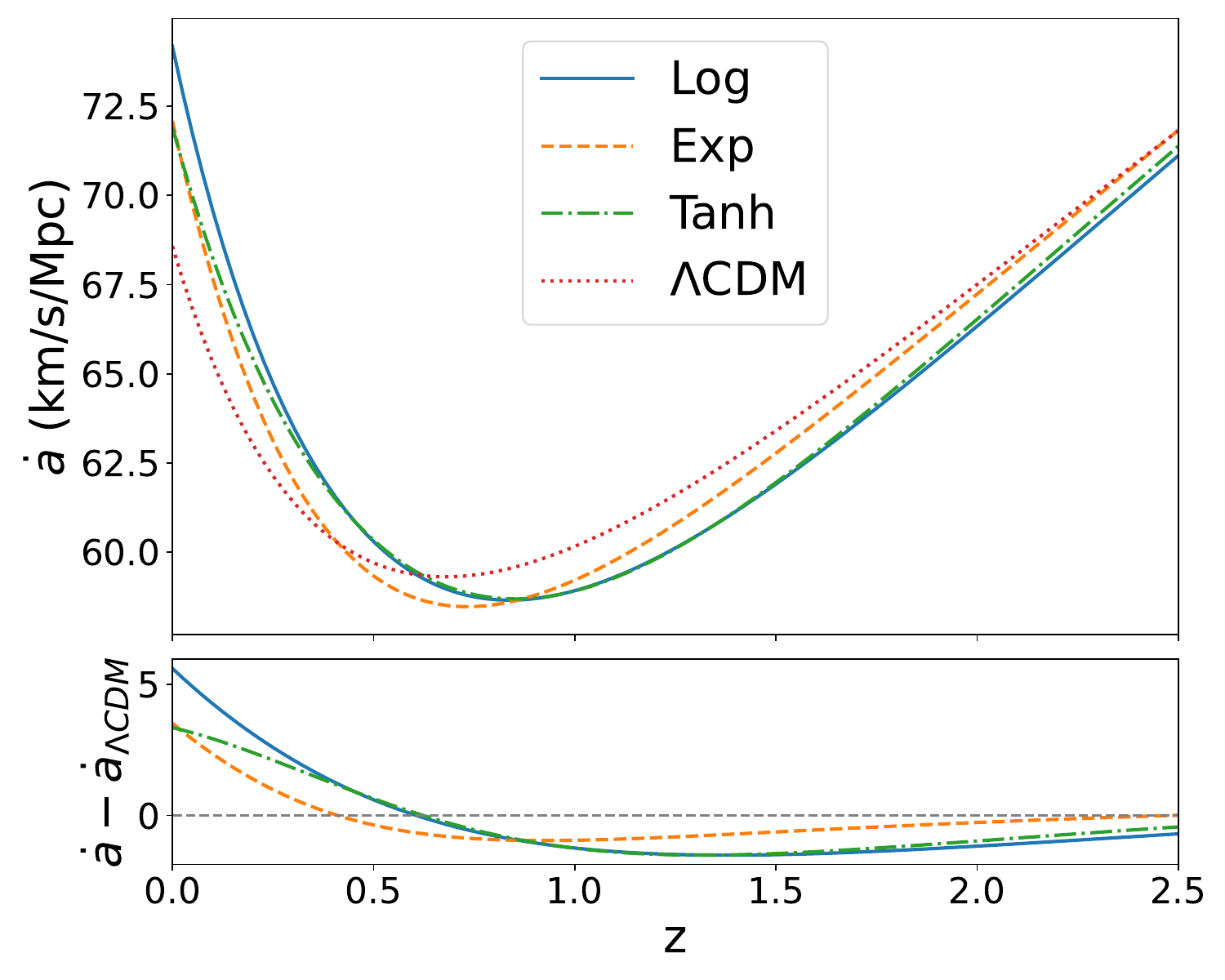}
\includegraphics[width=5.5cm]{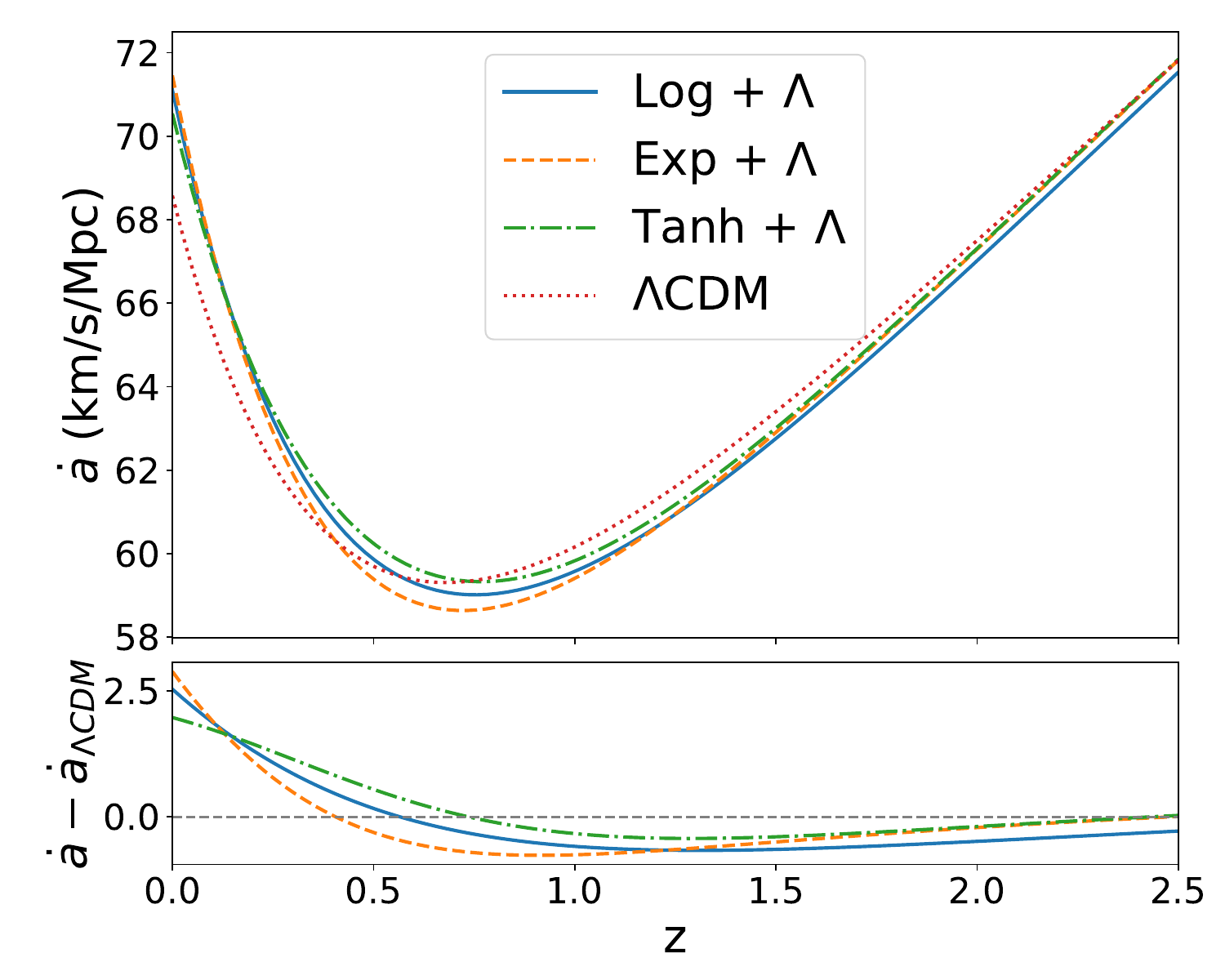}
\includegraphics[width=5.5cm]{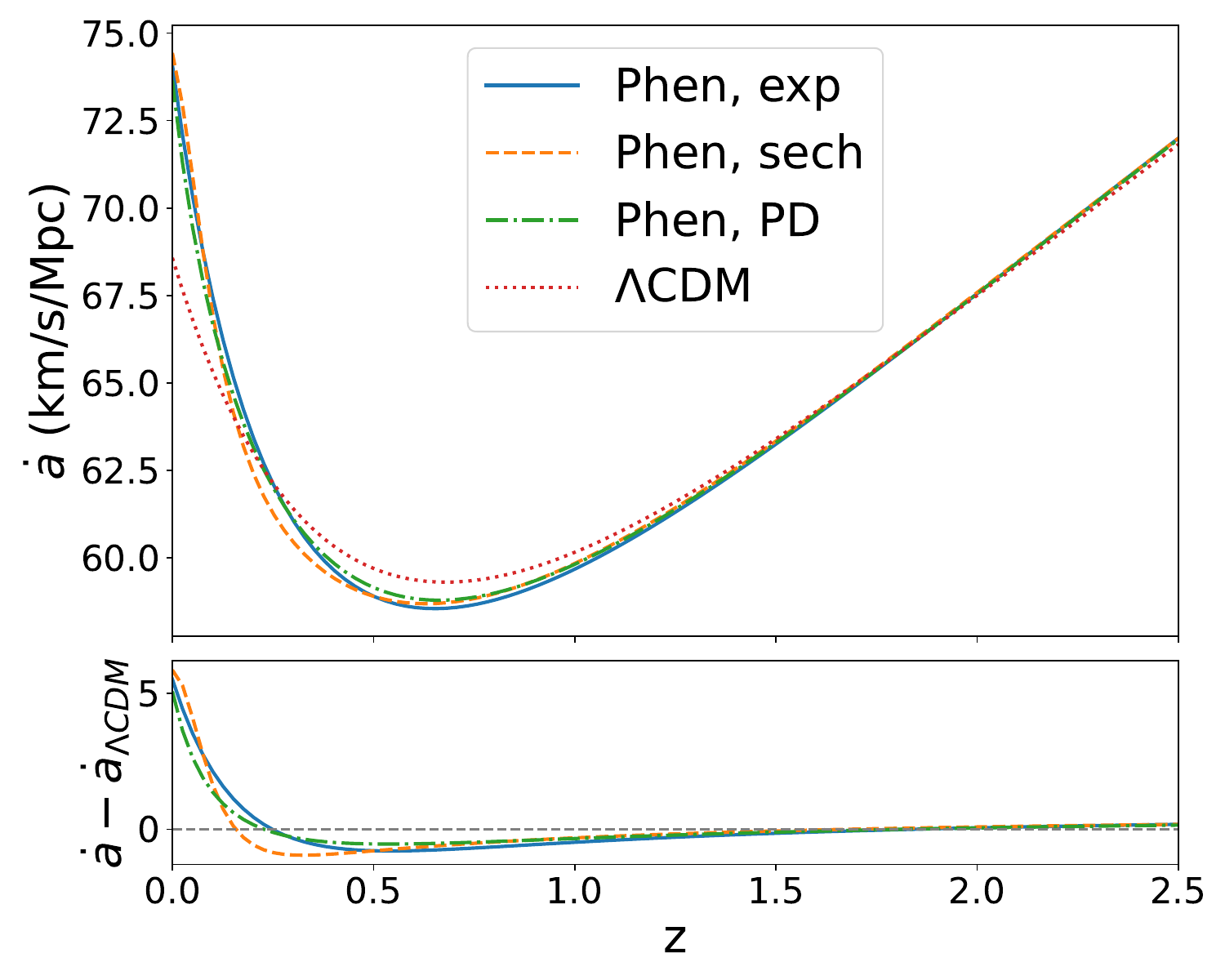}
\caption{Predicted $\dot{a}$ as a function of redshift for \textbf{(a)} the $f(Q)$ models, \textbf{(b)} the $f(Q)$ models assisted with the cosmological constant, and \textbf{(c)} the phenomenological models. The curves correspond to the mean $\dot{a}$ at each $z$ across MCMC samples, with the inferred cosmological parameters given in Tables~\ref{tab:ResultsfQ} and \ref{tab:ResultsPhen}. We also plot the $\Lambda$CDM case for comparison. Residuals with respect to it are shown in the lower panels. The $1\sigma$ C.L. error bars are not visible here due to the very tight CMB constraints.}
\label{fig:a_dot}
\end{figure}

In Figure~\ref{fig:a_dot}, we can see that all models have $\dot{a}(z)$ curves above $\Lambda$CDM at low redshift but below $\Lambda$CDM at higher redshift. This is entirely expected for models that aim to solve the Hubble tension without much altering the comoving distance to recombination. The transition occurs at a considerably lower redshift for the phenomenological models. The $f(Q)$ and Phen models depart considerably from $\Lambda$CDM at low $z$, but they converge to $\Lambda$CDM at high $z$. The deviations are milder in the $f(Q)+\Lambda$ models, limiting their ability to solve the Hubble tension.

\begin{figure}
\centering
\includegraphics[width=5.5cm]{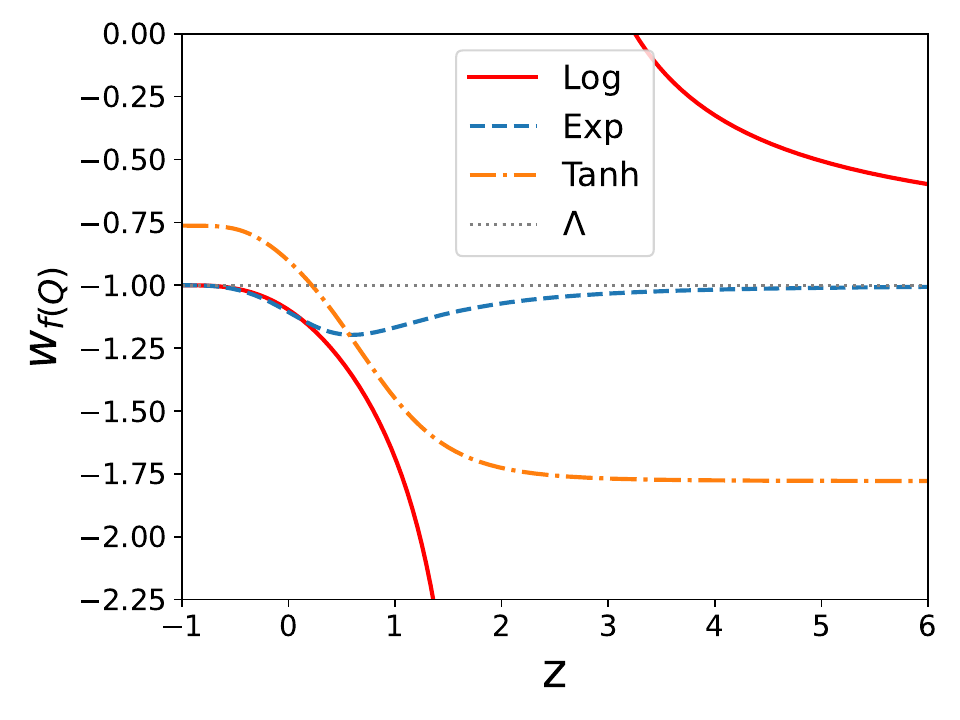} 
\includegraphics[width=5.5cm]{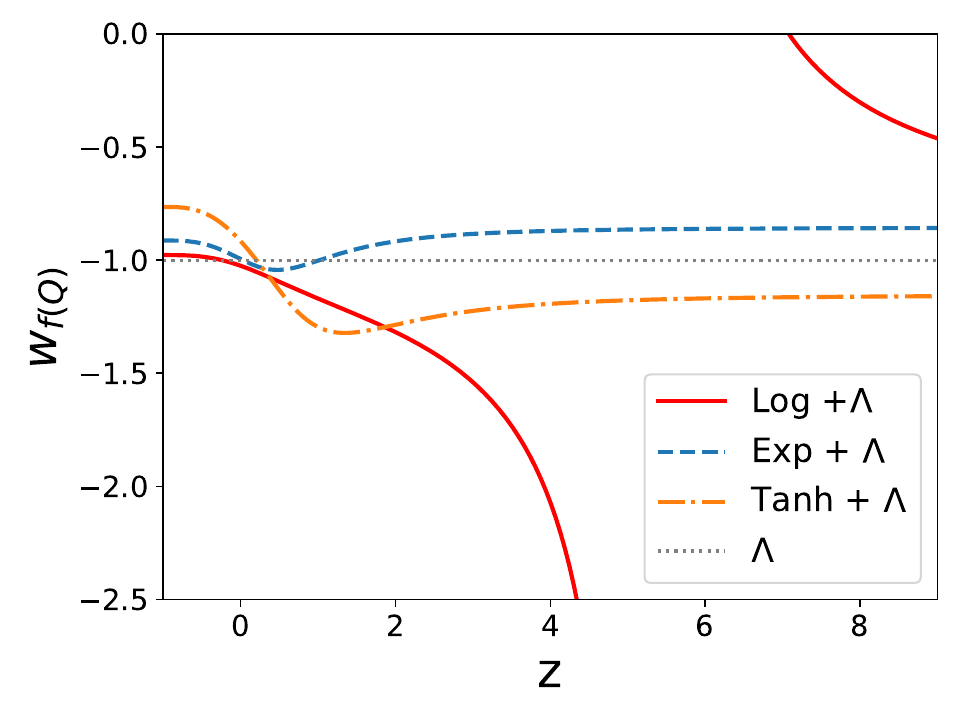}
\includegraphics[width=5.5cm]{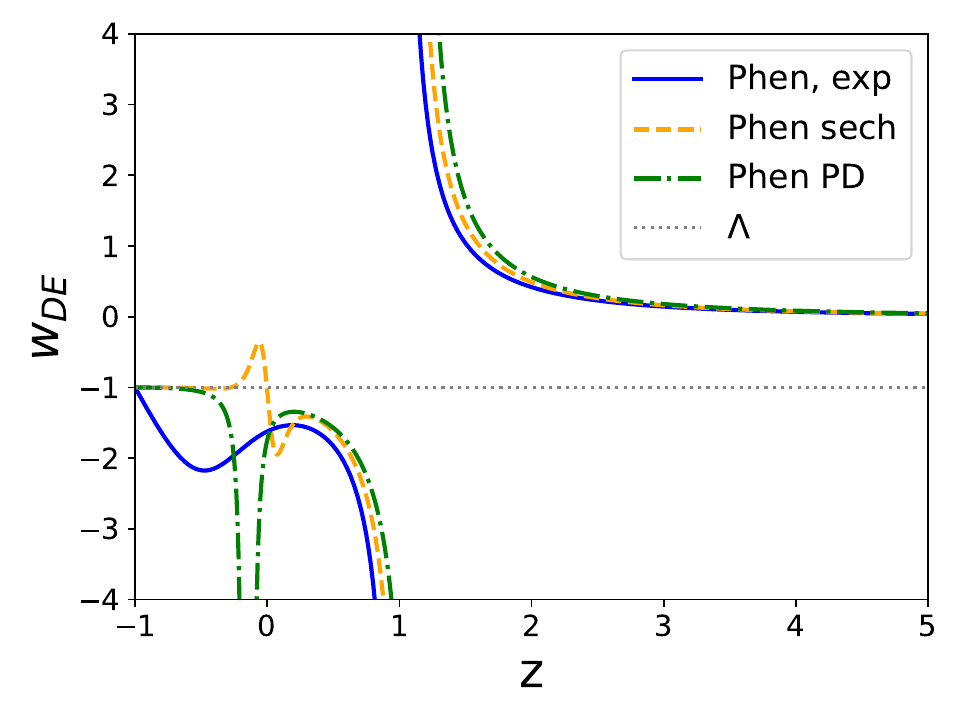}
\caption{The effective dark energy equation of state (EoS) of \textbf{(a)} the $f(Q)$ models, \textbf{(b)} the $f(Q)$ models assisted with the cosmological constant, and \textbf{(c)} the phenomenological models. The $1\sigma$ C.L. error bars are invisibly small due to the very tight \emph{Planck}~2018 constraints. 
}
\label{fig:w_EOS}
\end{figure}

Figure~\ref{fig:w_EOS} shows that the Exp model has a phantom behaviour that converges to the cosmological constant both at high redshift and in the infinite future ($z \to -1$). The Tanh model has an effective phantom crossing, with quintessence behaviour at very low redshift becoming phantom at higher redshift. The Log model has a singularity problem: it starts with phantom behaviour today, jumps to $w>-1/3$, and then becomes quintessence at high redshift. Close to $z=2$, the effective density changes sign, but the pressure stays negative. This creates a singularity in the effective equation of state ($P/\rho$). There is also a singularity for Log+$\Lambda$, but it happens at a higher redshift of around $z=6$. The Exp+$\Lambda$ and Tanh+$\Lambda$ models show a phantom crossing behaviour, with the former even having a double crossing. They converge to quintessence and phantom regimes at high redshift, respectively. Finally, they show quintessence behaviour when $z \to -1$. We stress that for the $f(Q)$ models, the EoS is an \textit{effective} EoS, meaning it does not come from a fluid but from treating the $f(Q)$ geometry as a fluid.

To obtain the dark energy EoS in the phenomenological models, we assume GR and apply the standard Friedmann equations. Figure~\ref{fig:w_EOS} shows that these models also have a singularity appearing around $z=1$. The Phen, sech model has a phantom crossing around $z=-0.1$. At high redshift, the model behaves like pressureless matter instead of dark energy. The Phen, PD model has a second singularity around $z=-0.15$. These future singularities show that the phenomenological models are physically unfeasible in GR.

Table~\ref{tab:JeffreysAgeResults} shows the statistical preference of the models relative to $\Lambda$CDM, assessed by the Bayesian Evidence. $\Lambda$CDM is very strongly preferred relative to the $f(Q)$ logarithmic model, but the remaining models are all very strongly preferred over $\Lambda$CDM on the Jeffreys scale. The models with the highest Bayes factors are the phenomenological models (particularly Phen, exp and Phen, sech) and the $f(Q)$ exp model.

\begin{table}
\caption{Goodness-of-fit comparison of the $f(Q)$ and phenomenological models with respect to the baseline $\Lambda$CDM model. 
A negative value for the log Bayes factor $B$ means that $\Lambda$CDM is preferred, whereas a positive value means a preference for the alternative $f(Q)$ or phenomenological model. We also present the posterior predicted age of the Universe for each model.}
\begin{tabular}{|c|c|c|}
\hline
\textbf{Model}	& $\ln B_{\text{model, $\Lambda$CDM}}$ & Age of the universe $t_0$ (Gyr)	\\
\hline
$\Lambda$CDM & - & $13.764 \, \pm \, 0.012$ \\
Log & $-12.736 \pm 0.228$ & $13.600 \, \pm \, 0.012$ \\
Log + $\Lambda$ & $14.939 \pm 0.234$ & $13.695 \, \pm \, 0.017$ \\
Exp & $16.637 \pm 0.226$ & $13.709 \, \pm \, 0.012$ \\
Exp + $\Lambda$ & $15.328 \pm 0.233$ & $13.719 \, \pm \, 0.014$ \\
Tanh & $5.744 \pm 0.228$ & $13.654 \, \pm \, 0.012$ \\
Tanh + $\Lambda$ & $12.031 \pm 0.233$ & $13.701 \, \pm \, 0.017$ \\
Phen, exp & $17.015 \pm 0.178$ & $13.740 \, \pm \, 0.015$ \\
Phen, sech & $17.015 \pm 0.179$ & $13.746 \pm 0.013$ \\
Phen, PD & $15.881 \pm 0.179$ & $13.745 \pm 0.012$ \\
\hline
\end{tabular}
\label{tab:JeffreysAgeResults}
\end{table}

Table~\ref{tab:JeffreysAgeResults} also shows that the predicted age of the Universe for all models is lower than for \emph{Planck}~2018 flat $\Lambda$CDM ($t_0 = 13.787 \, \pm \, 0.020$ Gyr). This is the case even for $\Lambda$CDM, which has $t_0 = 13.764 \, \pm \, 0.012$ Gyr. This is a direct consequence of the local constraints increasing $H_0$, since $t_0 \propto H_0^{-1}$.

\begin{table}
\caption{Minimum $\chi^2$ contributions for the considered models. We report separate contributions for the different observational constraints (DESI~DR2 BAO, local $H_0$, \emph{Planck}~2018 + SPT + ACT, and Cosmic Chronometers). We also show the number of data points in each catalogue.}
\begin{tabular}{|c|c|c|c|c|c|}
\hline
Data Points & 13 & 4 & 3 & 32 & 52 \\
\hline
\textbf{Model}	& $\chi^2_\text{DESI}$ & $\chi^2_{H_0}$ & $\chi^2_\text{CMB}$ & $\chi^2_\text{CC}$ & $\chi^2_\text{total}$	\\
\hline
$\Lambda$CDM & 10.88 & 41.89 & 13.44 & 14.84 & 81.06 \\
Log & 61.78 & 1.47 & 25.04 & 17.61 & 105.90 \\
Log + $\Lambda$ & 21.45 & 9.70 & 0.06 & 15.40 & 46.60 \\
Exp & 28.50 & 3.54 & 0.20 & 15.48 & 47.72 \\
Exp + $\Lambda$ & 24.33 & 6.20 & 0.90 & 15.29 & 46.73 \\
Tanh & 43.24 & 4.37 & 4.52 & 16.88 & 69.02 \\
Tanh + $\Lambda$ & 22.09 & 15.15 & 0.28 & 15.62 & 53.14 \\
Phen, exp & 19.24 & 2.54 & 3.77 & 14.94 & 40.48 \\
Phen, sech & 18.31 & 1.08 & 3.91 & 15.12 & 38.40 \\
Phen, PD & 19.27 & 2.95 & 26.65 & 14.84 & 63.71 \\
\hline
\end{tabular}
\label{tab:ChiSquaredResults}
\end{table}

\begin{figure}
\centering
\includegraphics[width=13.0cm]{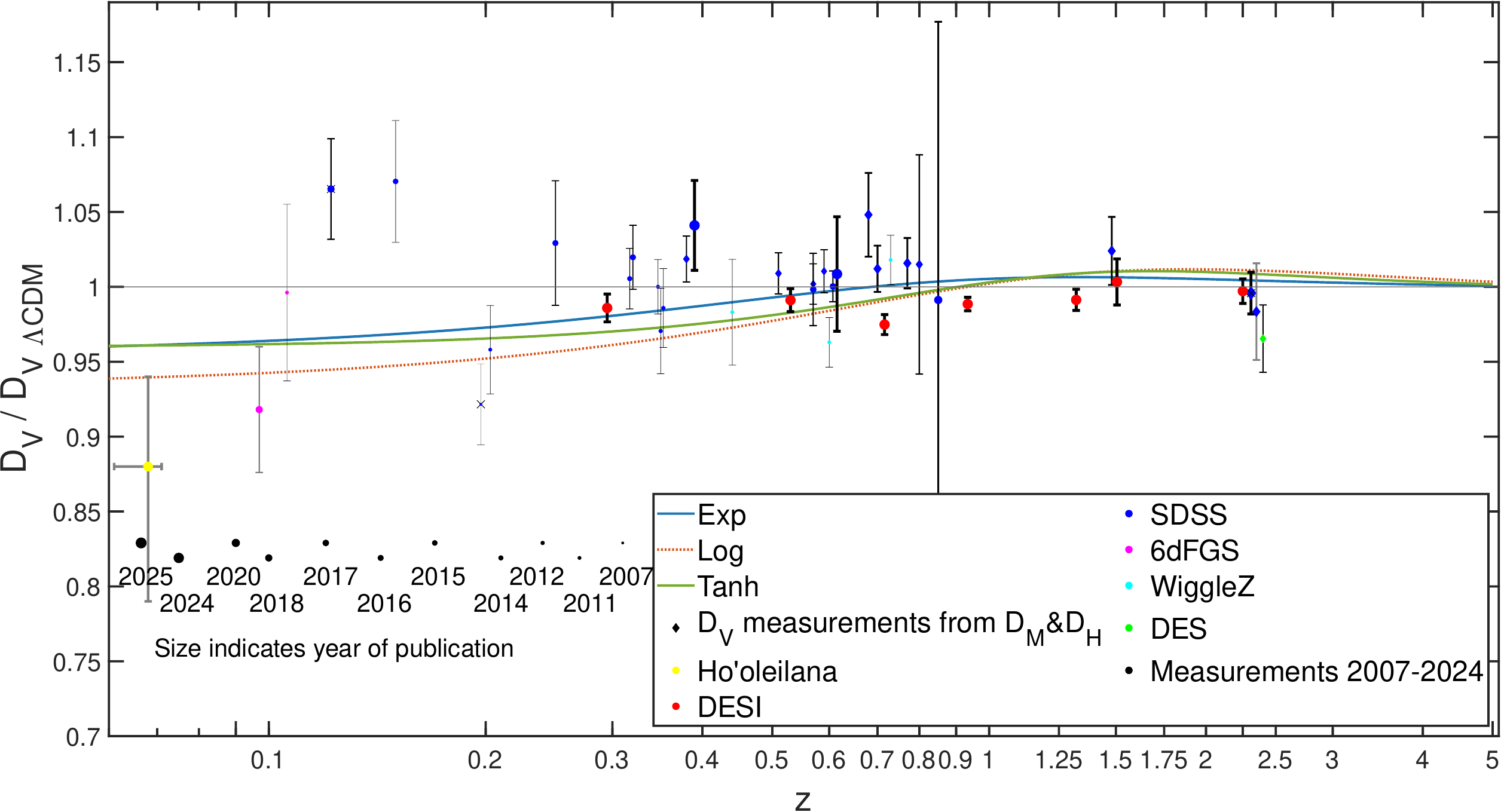} \\
\includegraphics[width=13.0cm]{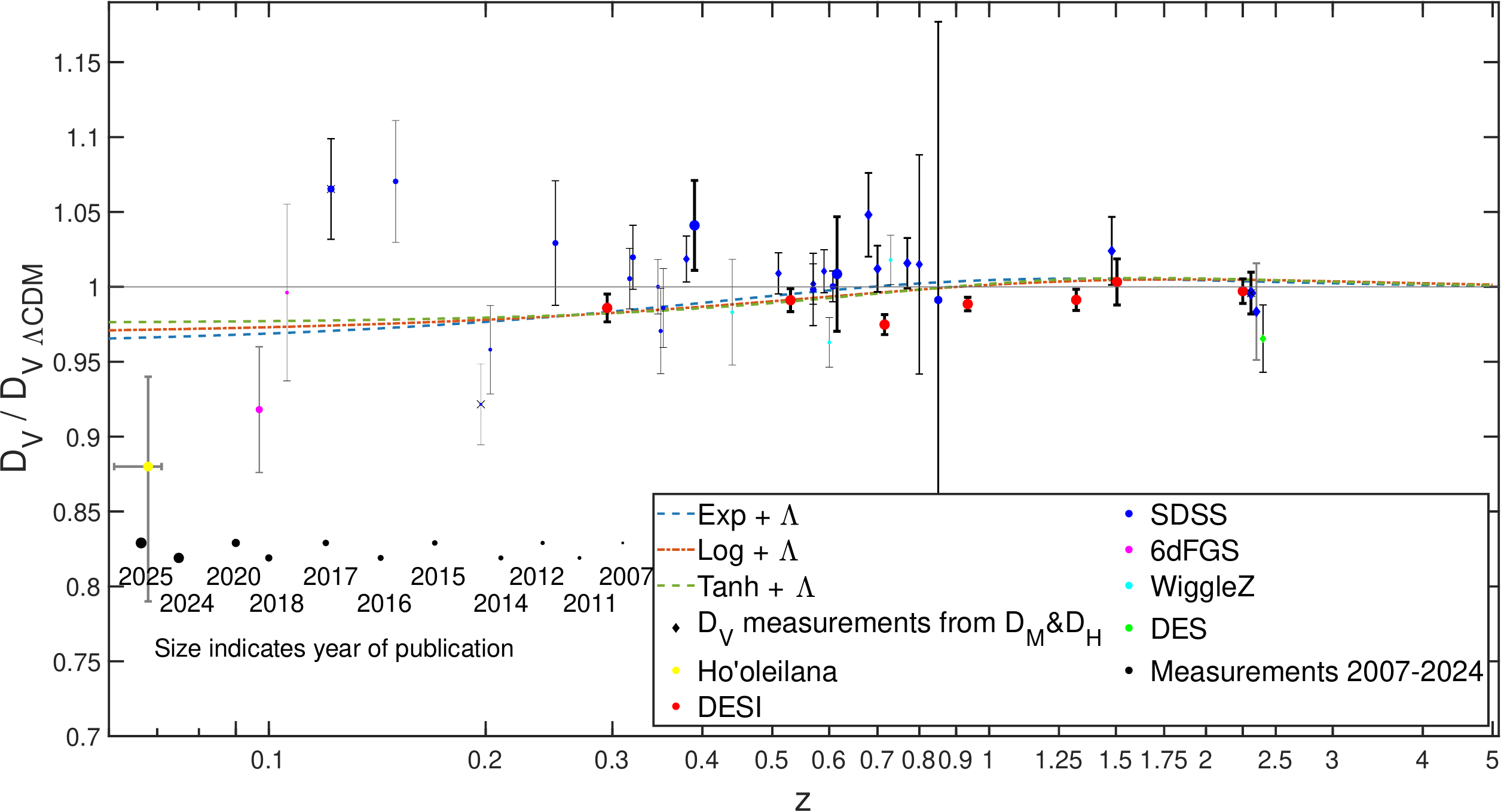}\\
\includegraphics[width=13.0cm]{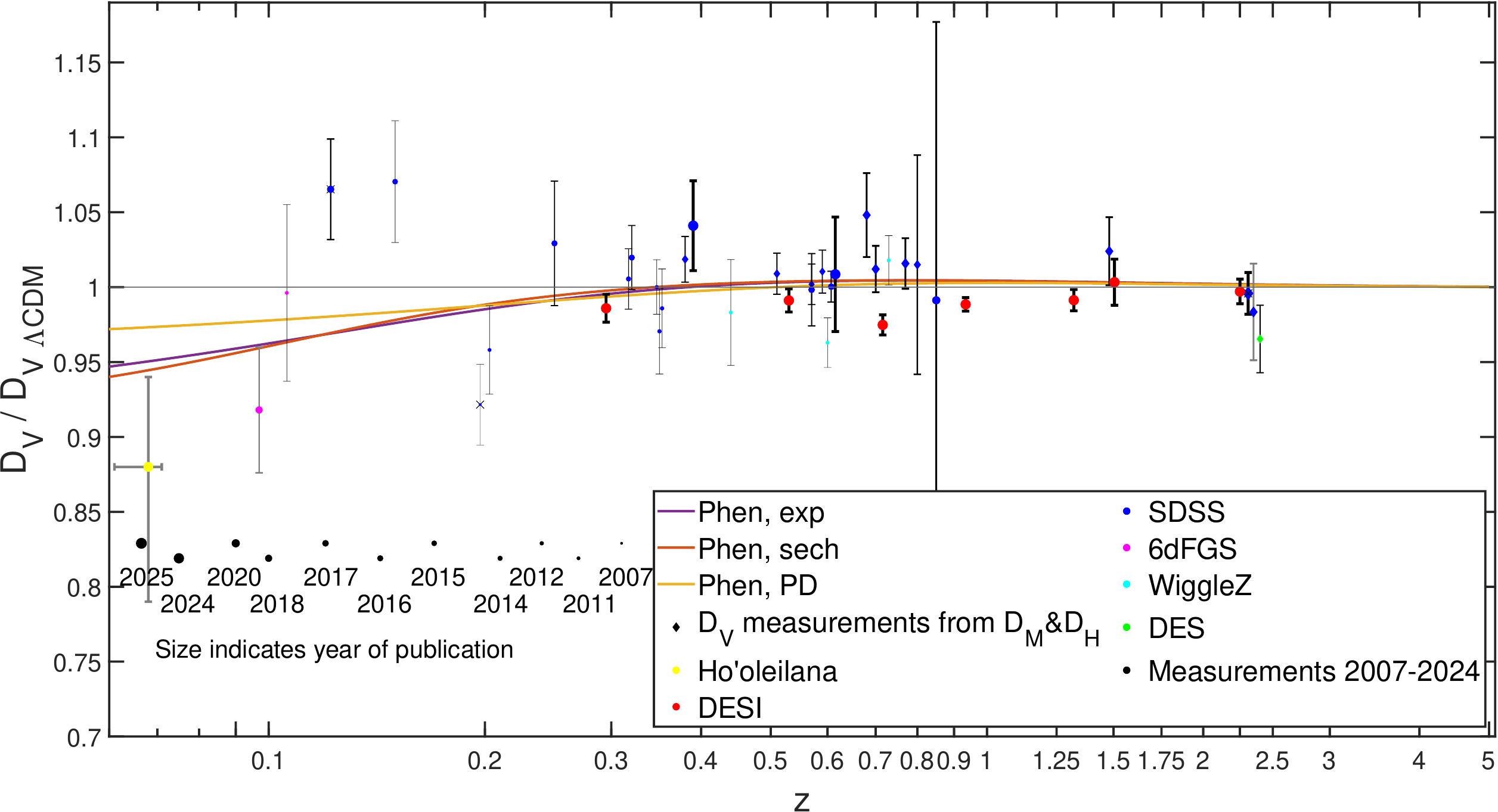}
\caption{Ratio between the isotropically averaged comoving BAO scale $D_V$ to the $\Lambda$CDM case for \textbf{(a)} the $f(Q)$ models, \textbf{(b)} the $f(Q)$ models plus cosmological constant, and \textbf{(c)} the phenomenological models. We take the best fits for all models with respect to the full dataset ($H_0$+\emph{Planck}~2018+SPT+ACT+CC+DESI~DR2 BAO), including the $\Lambda$CDM case. We also plot the BAO data from the last 20 years for comparison. 
}
\label{fig:DV_BAO}
\end{figure}

In Table~\ref{tab:ChiSquaredResults}, we can see that by far the best fit to DESI data is obtained by $\Lambda$CDM, even though it has the expected failure when confronted with the CMB and the reported local $H_0$. We will return to this BAO success later. $\Lambda$CDM calibrated using CMB alone gives $\chi^2_\text{DESI} \approx 31$ \citep{Banik_2025_BAO}. The $f(Q)+\Lambda$ and phenomenological models have noticeably better agreement with the DESI data. The $f(Q)$ models struggle to fit the BAO data, even the Exp model. We can corroborate this in Figures~\ref{fig:DV_BAO}, \ref{fig:DM_BAO}, and \ref{fig:DH_BAO}. This difficulty with the BAO data seems to explain why allowing a cosmological constant reduces the predicted $H_0$ to values below the local measurement (Figure~\ref{fig:f_Q_contours}).

Although $\Lambda$CDM struggles to fit the local $H_0$ determinations, the remaining models show a better fit to them, with the exception of the Tanh+$\Lambda$ model ($\chi^2_{H_0} > 15$ for 4 d.o.f.). This is due to these models predicting a higher $H_0$ of around 74~km/s/Mpc, while even the $f(Q) + \Lambda$ models predict around 71~km/s/Mpc. The CMB fit is reasonable for all models, except for $\Lambda$CDM, the logarithmic $f(Q)$ model, and the Phen, PD model. $\Lambda$CDM needs a higher $H_0$ to reduce the tension with the local $H_0$ measurements, decreasing the agreement with the CMB. Finally, all models overfit the CC data as $\chi^2\approx 15$, much lower than the number of CC data points (32). This suggests that the CC error bars might be overestimated \citep[as also found by][]{Kvint_2025}. Finally, the last column shows the total $\chi^2$. All models except for Log have a lower $\chi^2$ than $\Lambda$CDM, consistent with the Bayesian Evidence results. This does not necessarily mean the models are an improvement over $\Lambda$CDM, as the number of free parameters must also be considered.

Table~\ref{tab:ChiSquaredResults_BAO_CMB} shows the significance of the BAO anomaly in each model calibrated using the CMB data alone, either in the DESI data space or the $(\Omega_m, H_0 r_d)$ parameter space, with the MDPS significance highlighted. The Log and Tanh models have strong MDPS tensions ($7.55\sigma$ and $4.82\sigma$, respectively) from parameter space, while the remaining models are consistent within $2.7\sigma$. The Phen and $f(Q)+\Lambda$ models are consistent within $1.72\sigma$ according to the MDPS BAO-CMB tension, but none of these are theoretically well motivated. The $\Lambda$CDM and $f(Q)$ models are, but they show a considerable MDPS tension ($>2.6\sigma$, even reaching $7.55\sigma$ for Log and $4.82\sigma$ for Tanh). Interestingly, the Exp model has a slightly lower parameter space tension than $\Lambda$CDM, but the slightly higher data space tension means that the MDPS tension is the same for both models. This makes it the only theoretically well motivated alternative to $\Lambda$CDM which we find does not have a higher BAO tension than $\Lambda$CDM. However, the MDPS tension is not lower than $\Lambda$CDM either, indicating serious difficulties in reducing the MDPS tension that it faces when calibrated using CMB alone. Finally, we see that $\Lambda$CDM has a BAO tension in parameter space, but the tension is $1\sigma$ lower in data space. We discuss this further in Appendix~\ref{BAO_toy_model}.

\begin{table}
\caption{BAO significance from CMB-only fits. The MDPS tension is highlighted in bold on each row.}
\begin{tabular}{|c|c|c|c|c|}
\hline
\multirow{2}{2 cm}{\textbf{Model}} & \multicolumn{2}{c|}{DESI Data Space (13 d.o.f.)} & \multicolumn{2}{c|}{($\Omega_m, H_0 r_d$) Parameter Space} \\
\cline{2-5}
& $\chi^2_{DESI}$ & Gaussian tension ($\sigma$) & $\chi^2_{DESI}$ & Gaussian tension ($\sigma$) \\
\hline
$\Lambda$CDM & 19.93 & 1.66 & 9.65 & \textbf{2.65} \\
Log & 86.96 & 7.12 & 62.93 & \textbf{7.55} \\
Log + $\Lambda$ & 15.63 & \textbf{1.10} & 0.92 & 0.48 \\
Exp & 28.37 & \textbf{2.65} & 9.09 & 2.56 \\
Exp + $\Lambda$ & 18.47 & \textbf{1.47} & 0.07 & 0.04 \\
Tanh & 47.36 & 4.46 & 26.92 & \textbf{4.82} \\
Tanh + $\Lambda$ & 15.73 & \textbf{1.11} & 0.47 & 0.27 \\
Phen, Exp & 18.20 & \textbf{1.44} & 0.66 & 0.36 \\
Phen, Sech & 14.84 & 1.00 & 2.41 & \textbf{1.04} \\
Phen, PD & 20.44 & \textbf{1.72} & 3.32 & 1.31 \\
\hline
\end{tabular}
\label{tab:ChiSquaredResults_BAO_CMB}
\end{table}

The above results neglect local $H_0$ measurements. We therefore need to consider the significance of the BAO anomaly by constraining the model parameters using all the data apart from DESI (i.e., a DESI prediction from all non-BAO data) in the DESI data space and $(\Omega_m, H_0 r_d)$ parameter space. The major advantage here is that since the local $H_0$ is already included as a constraint, the models are now tuned to solve the Hubble tension as best as possible, with models other than $\Lambda$CDM achieving this quite well (Figure~\ref{fig:f_Q_contours}). This means that when comparing to BAO data, a low MDPS tension would indicate a good resolution to both the Hubble and BAO tensions. The other advantage is that models with additional degrees of freedom that arise primarily at late times cannot be properly constrained using the CMB alone. This can increase the uncertainty of the model predictions, artificially reducing the CMB-BAO tensions in Table~\ref{tab:ChiSquaredResults_BAO_CMB}.

Table~\ref{tab:ChiSquaredResults_BAO_noDESI} shows that the Log, Tanh, and Tanh+$\Lambda$ models have a strong MDPS BAO tension from parameter space ($6.82\sigma$, $4.93\sigma$, and $4.07\sigma$, respectively). These tensions cast very severe doubt on the models. The MDPS tension is $3.51\sigma$ for Log+$\Lambda$ and $3.18\sigma$ for Exp+$\Lambda$, which suggests they may be marginally viable, though still unlikely. The remaining models all come within the $2.7\sigma$ C.L. The theoretically well motivated Exp model has a moderate MDPS tension ($2.68\sigma$ C.L.). Although it performs relatively well among the considered models with reasonable theoretical justification, its MDPS tension is still higher than the $2.65\sigma$ MDPS tension faced by $\Lambda$CDM calibrated with CMB alone. Leaving aside the issue of theoretical motivation, the Phen models all come within $2.2\sigma$. The Phen, PD model has an MDPS tension of just $1.60\sigma$, but it cannot adequately fit the CMB (Table~\ref{tab:ChiSquaredResults}). This must be due to the much slower decay of the extra contribution to $H(z)$ (Eq.~\ref{eq:Delta_H_phen}), with the data requiring a faster decay. The Phen, exp and Phen, sech models can fit the CMB and local $H_0$, making them fairly promising in light of their low MDPS tension of only $2.13\sigma$ and $2.23\sigma$, respectively. This is driven by the data space tension in both cases. Although a $2.2\sigma$ tension is not obviously problematic, the bottom panel of Figure~\ref{fig:DV_BAO} shows that the Phen models do not really solve the BAO anomaly, deviating from \emph{Planck} in the opposite direction to the data at $z \approx 1$. These systematic residuals are not properly captured in parameter space, while in data space, the tension is somewhat diluted across 13 data points (see Appendix~\ref{BAO_toy_model}). Therefore, even our fairly flexible Phen models struggle to solve the BAO anomaly, as quantified by only a mild reduction in the MDPS tension compared to $\Lambda$CDM calibrated using CMB alone (the conventionally used measure of significance for the BAO anomaly).

\begin{table}
\caption{BAO significance from CMB+$H_0$+CC fits, i.e., using all non-BAO data to predict the BAO.}
\begin{tabular}{|c|c|c|c|c|}
\hline
\multirow{2}{2 cm}{\textbf{Model}} & \multicolumn{2}{c|}{DESI Data Space (13 d.o.f.)} & \multicolumn{2}{c|}{($\Omega_m, H_0 r_d$) Parameter Space} \\
\cline{2-5}
& $\chi^2_{DESI}$ & Gaussian tension ($\sigma$) & $\chi^2_{DESI}$ & Gaussian tension ($\sigma$) \\
\hline
$\Lambda$CDM & 11.20 & \textbf{0.53} & 0.72 & 0.39 \\
Log & 75.93 & 6.54 & 50.83 & \textbf{6.82} \\
Log + $\Lambda$ & 29.41 & 2.76 & 15.43 & \textbf{3.51} \\
Exp & 28.61 & \textbf{2.68} & 7.38 & 2.24 \\
Exp + $\Lambda$ & 25.39 & 2.32 & 13.04 & \textbf{3.18} \\
Tanh & 49.82 & 4.66 & 28.05 & \textbf{4.93} \\
Tanh + $\Lambda$ & 36.12 & 3.45 & 19.90 & \textbf{4.07} \\
Phen, Exp & 23.45 & \textbf{2.13} & 0.646 & 0.35 \\
Phen, Sech & 24.64 & \textbf{2.23} & 2.34 & 1.01 \\
Phen, PD & 19.43 & \textbf{1.60} & 3.95 & 1.48 \\
\hline
\end{tabular}
\label{tab:ChiSquaredResults_BAO_noDESI}
\end{table}

A very interesting aspect of Table~\ref{tab:ChiSquaredResults_BAO_noDESI} is that $\Lambda$CDM calibrated with all considered non-BAO datasets (CMB+$H_0$+CC) has perfect agreement with the BAO data, at $0.53\sigma$ C.L. This is a significant reduction from the $2.65\sigma$ tension it faces when calibrated using only the CMB. The $\chi^2$ in both data and parameter space drops by 9. Given the limited constraining power of CC datasets at present and the very precise local determinations of $H_0$ used in our analysis (section~\ref{sec:H0_estimates}), it is clear that the shifts the latter induce to the $\Lambda$CDM parameters substantially improve the fit to the DESI BAO measurements. This seems a rather unlikely coincidence if the precise local $H_0$ measurements are affected by unknown systematics, since then the inferred $\Lambda$CDM parameters using CMB+$H_0$+CC would also be inaccurate, presumably feeding through to inaccurate yet precise BAO predictions. These cannot generally be expected to line up with the actual BAO measurements, which are affected by completely different systematics to local estimates of $H_0$. Our interpretation is that although $\Lambda$CDM cannot simultaneously fit the CMB and the locally reported $H_0$, the best fit to these datasets is a good approximation to the true expansion history, especially at the intermediate epochs probed by BAO measurements. In short, using the local $H_0$ determinations as an additional constraint on the $\Lambda$CDM model corrects it to some extent for new physics at low $z$, reducing the BAO tension to a negligible level.

\section{Discussion}
\label{sec:Discussion}

The first result to highlight from this paper is that it is possible to solve the Hubble tension at the background level without early-time new physics (in line with the recent results from \cite{Jia_2025b, Lopez_2025, Kavya_2025}). We can see this from the $H_0$ constraints and the $\chi^2$ results for the $H_0$ and CMB datasets, which can be fit simultaneously in most models. The only models that struggle to fit the CMB constraints are $\Lambda$CDM, the $f(Q)$ Log model, and the Phen, PD model, with a moderate Hubble tension also faced by the Tanh+$\Lambda$ model (Table~\ref{tab:ChiSquaredResults}). Since $\Lambda$CDM is usually calibrated using the CMB and is known to provide an excellent fit to it \emph{alone}, the tension here is caused by requiring a simultaneous fit to the low-redshift $H_0$ measurements. By raising $H_0$ to try and fit them, the agreement with the CMB is slightly worsened ($>2\sigma$ tension). The $\Lambda$CDM model is then formally in tension with both the $H_0$ and CMB measurements, recovering the well-known Hubble tension.

The next important result is that the $f(Q)$ models calibrated using the CMB perform worse than $\Lambda$CDM when it comes to predicting the DESI~DR2 data, apart from the Exp model which has the same level of tension as $\Lambda$CDM (Table~\ref{tab:ChiSquaredResults_BAO_CMB}). We can also see a trend for the $f(Q)$ models when adding the cosmological constant $\Lambda$, which significantly reduces $\chi_{\mathrm{DESI}}^2$ in both data and parameter space. This reduces the overall $\chi^2$, despite the larger $\chi_{H_0}^2$ contribution. We can visually see the BAO tensions of these models in Figures~\ref{fig:DV_BAO}, \ref{fig:DM_BAO}, and \ref{fig:DH_BAO}. Although adding $\Lambda$ considerably improves the ability of the $f(Q)$ models to predict the DESI data using CMB alone, their theoretical motivation is undermined as alternatives to a cosmological constant. Even so, these hybrid $f(Q)+\Lambda$ models are useful examples of how a background solution to the CMB-BAO tension could work, neglecting the issue of the Hubble tension.

This is taken into account in Table~\ref{tab:ChiSquaredResults_BAO_noDESI}, where we show the DESI data predictions using CMB+$H_0$+CC. Compared to the CMB-only case, the $f(Q)$ models show similar results in both data and parameter space, indicating that the models could already solve the Hubble tension without being forced to do so. The $f(Q)+\Lambda$ models have considerably higher tensions in the CMB+$H_0$+CC case. This is mainly caused by the fact that the extra $\Lambda$ term introduces an additional degree of freedom at late times, meaning the CMB alone cannot constrain the model parameters as well as in the $f(Q)$ models, which have the same number of free parameters as $\Lambda$CDM. The inability of the CMB alone to constrain the model parameters leads to uncertain BAO predictions, artificially reducing the tension with DESI. This problem can be rectified by adding the $H_0$ data, which strongly constrains deviations from $\Lambda$CDM at late times. Thus, the higher $\chi_{\mathrm{DESI}}^2$ when using CMB+$H_0$+CC is much more realistic.

Although the Phen models are not theoretically motivated (unlike the $f(Q)$ models), they generally outperform $\Lambda$CDM in its ability to predict the BAO data using the CMB alone (Table~\ref{tab:ChiSquaredResults_BAO_CMB}). However, this comes from the Phen models having two extra parameters beyond $\Lambda$CDM that only have an effect at very low $z$, leading to imprecise parameter constraints using the CMB alone. This increases the uncertainty of the BAO predictions, thereby decreasing $\chi^2_{\mathrm{DESI}}$. This issue is ameliorated by introducing $H_0$ measurements as extra constraints and predicting the BAO data using CMB+$H_0$+CC. The Phen, exp and Phen, sech models now have mild MDPS tensions with the BAO data ($2.13\sigma$ and $2.23\sigma$ C.L., respectively), both from data space. Even the latter is still below the MDPS tension faced by $\Lambda$CDM calibrated using CMB alone, which is the conventional measure of the BAO anomaly. Table~\ref{tab:JeffreysAgeResults} also shows that these models have a Bayesian Evidence much higher than $\Lambda$CDM, so that empirically speaking, they provide a compelling solution to the Hubble and BAO tensions. Indeed, there is hardly any parameter space BAO tension to speak of. Figures~\ref{fig:DV_BAO}, \ref{fig:DM_BAO}, and \ref{fig:DH_BAO} illustrate that the phenomenological models are in reasonable agreement with the great majority of BAO data accumulated over the last 20 years. However, the Phen models systematically overpredict isotropically averaged BAO distances from DESI~DR2 at $z \lesssim 1.5$ (Figure~\ref{fig:DV_BAO}). This leads to a particularly large discrepancy at $z = 0.7$, which must be revisited once future data becomes available. We also note that any genuine BAO tension in the Phen models would be diluted by the still somewhat uncertain model parameters when using all non-BAO datasets (Figure~\ref{fig:phen_contours}).

Another highlight for the Phen models is that $z_1 \lesssim 0.2$. This suggests that to solve the Hubble tension consistently with BAO data, we need an extra acceleration term beyond $\Lambda$CDM that only arises at very low redshift \citep[close to the $z<0.15$ regime of][]{Riess_2022_comprehensive}, hinting that the solution to both problems lies just beyond the third rung of the distance ladder. Importantly, the inferred $z_1 \ll z_\Lambda$, where $z_\Lambda \approx 0.5$ is the redshift at which $\ddot{a}$ switches sign due to dark energy or something with a similar effect. If the observed accelerated expansion of the Universe is not due to $\Lambda$ but instead arises from some other cause like a geometric effect, we would expect our Phen models to indicate that departures from $\Lambda$CDM set in when $\Lambda$ becomes dominant, i.e., that $z_1 \approx z_\Lambda$. But our inferred $z_1 \ll z_\Lambda$, indicating that the deviation from $\Lambda$CDM identified by the Phen models and necessary to solve the Hubble and BAO tensions is probably not related to alternative interpretations of the accelerated cosmic expansion.

While a departure from the $\Lambda$CDM expansion history at such low $z$ is possible, the low inferred $z_1$ in the phenomenological models might instead be interpreted to mean that the solution is local in nature. This is possible if we live in a large local underdensity or void, as suggested by galaxy number counts \citep{Keenan_2013}. A model set up along those lines to solve the Hubble tension \citep{Haslbauer_2020} correctly predicted the BAO anomaly \citep{Banik_2025_BAO} and an apparent descending trend in the value of $H_0$ with the redshift of the data used \citep{Mazurenko_2025, Jia_2025b}. The data space tension in all three considered void models is lower than for our Phen, exp and Phen, sech models, the only Phen models consistent with the CMB \citep[table~3 of][]{Banik_2025_BAO}. This is true despite those authors neglecting uncertainties on the \emph{Planck} 2018 parameters and thus on the model predictions, which are based on a previously published analysis that did not consider BAO \citep{Haslbauer_2020}. The reason for this success is that the void models predict an $\alpha_\text{iso}(z)$ curve that nicely passes through the DESI~DR2 measurements \citep[figure~4 of][]{Banik_2025_BAO}. For a recent review of the local void solution, we refer the reader to \cite{Banik_2026_void}. The main difference with our approach is that the local void solution involves modifications to $\Lambda$CDM at the perturbation level with a background \emph{Planck} cosmology, in contrast to the purely background solutions considered here that retain the assumption of purely cosmological redshift.

The Bayesian Evidence shows that the Log $f(Q)$ model provides a much worse fit than standard $\Lambda$CDM, so it cannot be considered a candidate to solve the Hubble tension while achieving BAO consistency. However, adding the cosmological constant allows the model to outperform $\Lambda$CDM, with very strong Evidence. This comes at the cost of undermining its theoretical motivation, which was to explain the accelerated expansion of the Universe without dark energy \citep{Najera_2023}. Adding the cosmological constant reduces the effective geometric density to $\Omega_\alpha = 0.381 \, \pm \, 0.053$ for the Log model. Thus, the hybrid model can only explain around a third of the universe's energy content, or half the canonical dark energy contribution. It also cannot predict the DESI data using all non-DESI datasets (Table~\ref{tab:ChiSquaredResults_BAO_noDESI}). 

Even without a cosmological constant, the Exp and Tanh $f(Q)$ models have very strong preference against the flat $\Lambda$CDM model on the Jeffreys scale. This puts them in a good position to solve the Hubble tension with reasonable theoretical motivation. Unfortunately, Tables~\ref{tab:ChiSquaredResults} and \ref{tab:ChiSquaredResults_BAO_CMB} show that they are not fully consistent with BAO. In particular, the MDPS tension with BAO is higher than or at best equal to the $2.65\sigma$ C.L. tension for $\Lambda$CDM in the traditional $(\Omega_m, H_0 r_d)$ parameter space quantification of the BAO tension \citep{DESI_2025, Camphuis_2025}. The level of BAO tension for the Exp and Tanh $f(Q)$ models is quite similar in data and parameter space. The MDPS tensions also remain similar if the model parameters are instead calibrated using CMB+$H_0$+CC, highlighting that the models can naturally account for the high local $H_0$ and do not `shift' much when its observed value is added as a constraint. The lowest MDPS tension of the considered $f(Q)$ models fit to CMB+$H_0$+CC is $2.68\sigma$ for the Exp model, making it the strongest theoretically motivated model considered in this paper. The inability of any of the considered $f(Q)$ models not assisted by $\Lambda$ to reduce the MDPS tension faced by $\Lambda$CDM is a significant result, highlighting the difficulty of solving the Hubble tension consistently with BAO data. For the $f(Q)$ models with $\Lambda$, the same conclusion applies, provided the models are constrained using CMB+$H_0$+CC to ensure that the local $H_0$ is used. Including $\Lambda$ as an extra degree of freedom at late times makes it difficult to accurately constrain the model parameters using CMB alone, so it is important to consider the local $H_0$ measurements. The MDPS tensions of the $f(Q)+\Lambda$ models fit to CMB+$H_0$+CC are all $\geq 3.18\sigma$, making them considerably worse than $\Lambda$CDM in the usual way of quantifying its BAO tension.

By looking at Tables~\ref{tab:ResultsfQ}, \ref{tab:ResultsPhen}, \ref{tab:ChiSquaredResults_BAO_CMB}, and \ref{tab:ChiSquaredResults_BAO_noDESI}, we see that it is difficult to obtain a joint background solution to the Hubble and BAO tensions. While the $f(Q)$ and most Phen models solve the Hubble tension, they struggle to predict the BAO data, as quantified by the MDPS tension. In Table~\ref{tab:ChiSquaredResults_BAO_noDESI}, it seems the Phen, PD model solves the Hubble and BAO tensions. However, in Table~\ref{tab:ChiSquaredResults}, we see that the model has a strong tension with the CMB data. A full solution to the Hubble tension necessarily implies having a high $H_0$ while being consistent with the CMB. Therefore, the Phen, PD model cannot solve the CMB-$H_0$ or Hubble tension. The remaining models solve the Hubble tension, but the MDPS tension with the BAO data is $\geq 2\sigma$ if predictions are based on CMB+$H_0$+CC. This is important because otherwise we do not include $H_0$, and also because many of our models have extra free parameters that primarily affect the expansion history at low $z$, preventing the CMB alone from accurately constraining the model parameters. Thus, quantifying the BAO tension using DESI predictions based on CMB alone is not that realistic. Doing so would lead to uncertain model predictions, artificially reducing $\chi^2_{\mathrm{DESI}}$. This issue seems to be particularly severe for the $f(Q)+\Lambda$ models, which seem to be consistent with BAO data in the CMB-alone case, but not when using CMB+$H_0$+CC. When constrained in this way, none of the considered models reduce the MDPS tension with DESI to within the $2\sigma$ C.L., implying an at best marginal reduction to the MDPS tension of $\Lambda$CDM using CMB alone. While the $2.1\sigma-2.2\sigma$ tension achieved by the Phen models is not obviously a problem, a closer look shows that these models systematically overpredict $D_V/r_d$ compared to the DESI~DR2 measurements at all redshifts (Figure~\ref{fig:DV_BAO}). We note that since the Phen models have two extra free parameters compared to $\Lambda$CDM (Eq.~\ref{eq:Delta_H_phen}), the parameters are not that well constrained (Figure~\ref{fig:phen_contours}), leading to somewhat uncertain BAO predictions. This has presumably reduced the data space tension with the DESI observations.

When $\Lambda$CDM is calibrated using the CMB, there is no significant BAO tension in data space ($1.66\sigma$ C.L.), but there is a moderate tension in the $(\Omega_m, H_0 r_d)$ parameter space \citep[as previously reported in][]{Camphuis_2025, DESI_2025}. This means $\Lambda$CDM predicts the BAO data without significant tension, but it has a tension in the predicted cosmological parameters. This is indicative of a systematic trend in the residuals between predictions and observations. In Appendix~\ref{BAO_toy_model}, we use a toy model to illustrate that a $1.66\sigma$ data space tension and a $2.65\sigma$ parameter space tension are quite unlikely to arise simultaneously if the underlying model is correct, but if there are systematic deviations from it, then this combination is far more plausible. A systematic deviation from $\Lambda$CDM at low redshifts is also suggested by the preference for a negative neutrino mass sum \citep{DESI_2025_neutrinos}. Since neutrinos are ultra-relativistic at high $z$, their rest mass is only relevant at low $z$, making the inferred neutrino mass sum a diagnostic for anomalous behaviour at low $z$.

Another strong case for departures from $\Lambda$CDM is the Hubble tension, which also arises at low $z$ \citep{Lin_2021, Perivolaropoulos_2024, Banik_2025_cosmology, Jia_2025b, Lopez_2025, Banik_2026_void, Pantos_2026}. As with the other models, we could take it into account by constraining the $\Lambda$CDM model parameters using CMB+$H_0$+CC when predicting BAO data. Table~\ref{tab:ChiSquaredResults_BAO_noDESI} shows agreement in data (parameter) space at the $0.53\sigma$ ($0.39\sigma$) C.L., demonstrating that the BAO anomaly drops to negligible levels if we include $H_0$ constraints. This is surprising given that $\Lambda$CDM has a strong Hubble tension, which is evident in Table~\ref{tab:ChiSquaredResults} from the high $\chi^2_{\mathrm{CMB}}$ and $\chi^2_{H_0}$ when doing a simultaneous fit to the CMB and $H_0$. However, the very low $\chi^2_{\mathrm{DESI}}$ of 10.9 for 13 data points shows that the BAO anomaly `disappears'. This is somewhat misleading because $\Lambda$CDM cannot jointly fit the datasets used to constrain its model parameters when computing these BAO predictions. Even so, it is interesting that the BAO anomaly can be removed simply by constraining $\Lambda$CDM using CMB+$H_0$+CC instead of CMB alone. If the local $H_0$ measurements were inaccurate, given they are very precise, they would shift the $\Lambda$CDM parameters to incorrect values from those inferred using the CMB alone. The resulting incorrect parameters would typically be expected to yield incorrect BAO predictions. In fact, the BAO predictions obtained in this way are remarkably close to the observations, far more so than if fitting $\Lambda$CDM to the CMB alone. We suggest that this is unlikely to arise by chance. Our interpretation is instead that both the CMB and local $H_0$ measurements are correct, so fitting $\Lambda$CDM to both constraints provides a good approximation to the actual expansion history -- especially at the intermediate epochs probed by DESI.

In Table~\ref{tab:JeffreysAgeResults}, we can see that raising $H_0$ slightly reduces the age of the universe $t_0$ in the models.
\begin{equation}
    t_0 = \frac{1}{H_0} \int^\infty_0 \frac{a}{E(z)} \, dz,
\end{equation} 
where $E(a) \equiv H(a)/H_0$ is the reduced Hubble factor. The comoving distance to recombination is given by a similar integral, but with $c$ in the numerator instead of $a$ (Eq.~\ref{eqn:comovingDistRec}). To preserve the CMB distance to recombination and be consistent with the local $H_0$, we need to decrease $1/H(z)$ at low redshift and compensate by increasing it at high redshift to produce the same integral. However, the $t_0$ integral has an extra factor of $a$ in the numerator, making changes at low $z$ relatively more important. This means that decreasing $1/H(z)$ at low $z$ has a bigger impact, since $a \approx 1$ in this range. At high $z$, increasing $1/H(z)$ has the opposite effect, but high redshifts play a smaller role in $t_0$ because these are downweighted by the small $a$. This means that raising $H_0$ to match the local measurements while preserving the comoving distance to recombination has an overall effect of reducing the age of the universe. Consequently, all the models predict lower ages for the Universe. These ages are in the range $13.60-13.74$~Gyr with small uncertainties, since we use the tight CMB constraints \citep{Planck_2020, Camphuis_2025, Louis_2025}. The extent to which these ages are possible can be assessed using the ages of the oldest Galactic stars \citep{Cimatti_2023, Lundkvist_2025} and globular clusters \citep{Valcin_2025}. There are stars with ages higher than $13.3$ Gyr \citep{Cimatti_2023}, so the Universe must be older than this age. The samples considered in their study have an inverse variance weighted mean age of $14.05 \pm 0.25$~Gyr, implying that $t_0 \approx 14.25 \pm 0.25$~Gyr if the objects they studied took 200~Myr to form. The models considered in this paper are therefore in $\approx 2\sigma-3\sigma$ tension with their estimated age of the Universe. Indeed, those authors found that $H_0 < 73$ km/s/Mpc (assuming $\Lambda$CDM) with a probability of 90.3\% \citep{Cimatti_2023}. Furthermore, globular clusters imply $t_0 = 13.57^{+0.16}_{-0.14} \, \text{(stat)} \, \pm \, 0.23 \, \text{(sys)}$ Gyr \citep{Valcin_2025}. Since those authors report that the oldest globular cluster has an age of 13.39~Gyr, this implicitly assumes a formation time of only 180~Myr. Assuming this is valid, all the considered models studied in this paper lie within the $1\sigma$ C.L. of the age reported by \citep{Valcin_2025}. Future studies will shed more light on the feasibility of lower ages of the Universe than \emph{Planck} $\Lambda$CDM, which is already in slightly under $2\sigma$ tension with the higher age reported in \cite{Cimatti_2023, Lundkvist_2025}. A reduced age is a necessary feature for background late-time solutions to the Hubble tension. This is also a feature of models that solve only the BAO anomaly without considering the Hubble tension, since the BAO anomaly implies a smaller comoving distance to $z \approx 1$ than \emph{Planck} $\Lambda$CDM \citep{Banik_2025_BAO}. For instance, the CPL $w_0w_a$ parametrization implies an age of $t_0 = 13.76$ Gyr (using DESI~DR2+CMB data), higher than the models considered in this paper but lower than \emph{Planck} $\Lambda$CDM.

Figure~\ref{fig:w_EOS} highlights several problems that some of the models have in the effective dark energy EoS, where the Log and Log+$\Lambda$ models have a singularity. This comes from the effective density changing sign while the effective pressure remains negative. Since this is an effective EoS, there is no singularity in the true EoS because the universe in this cosmology is composed only of matter and radiation. For the phenomenological models, this is not the case as it is a true EoS coming from dark energy. The singularity in them is a true pathology (the Phen, PD has two singularities). Therefore, the phenomenological models imply an unphysical energy density, violating the weak energy condition. While this suggests the models are not physically viable as a simple dark energy fluid in GR, it might instead point towards new physics in the form of modified theories of gravity. They might accommodate a solution like Eq.~\ref{eqn>H_z_phen} which can solve the observational tensions studied in this paper. The remaining models do not have singularity problems. This includes the $f(Q)$ Exp model, which is the strongest model overall since it alleviates the Hubble and BAO tensions in the $(\Omega_m, H_0 r_d)$ parameter space, its Bayes factor is high (16.637, just short of the 17 reached by the Phen models), and it is theoretically motivated. The Exp model shows a phantom behaviour converging to the cosmological constant at high redshift and also as $z \to -1$, which represents the infinite future. This is crucial as phantom behaviour when $z \to -1$ implies a Big Rip. The Exp + $\Lambda$ model has two phantom crossings, one close to the present time and one at $z \approx 1$. Inside that range, the behaviour is phantom, whereas outside it is quintessence. The Tanh and Tanh + $\Lambda$ models have phantom behaviour at high redshift and exhibit a phantom crossing close to the present time, having quintessence behaviour from now to the infinite future. While this avoids a Big Rip singularity, the models are among the poorest performers according to the overall Bayesian Evidence, albeit still better than $\Lambda$CDM (Table~\ref{tab:JeffreysAgeResults}).

\section{Conclusions}
\label{sec:Conclusions}

We study modifications to the $\Lambda$CDM background evolution in the context of the Hubble and BAO tensions. We focus on three $f(Q)$ modified gravity models and three phenomenological models that supplement the $\Lambda$CDM $H(z)$ with an exponential (Phen, exp), hyperbolic secant (Phen, sech), and polynomial decay (Phen, PD) term at low $z$ (Eq.~\ref{eqn>H_z_phen}). The theoretical models are given by $f(Q) = Q/(8 \pi G) - \alpha \ln(Q/Q_0)$ (``Log''), $f(Q) = Q/(8\pi G) \times \exp(\lambda Q_0/Q)$ (``Exp''), and $f(Q) = Q/(8 \pi G) + \alpha \tanh (Q_0/Q)$ (``Tanh''). We also consider the same models assisted with the cosmological constant $(``+\Lambda'')$.

We constrained the cosmological parameters with four different datasets: local $H_0$ measurements, DESI~DR2 BAO, CMB constraints on $(100 \, \theta_\star, w_b, w_{bc})$ with \emph{Planck}~2018+SPT-3G+ACT~DR6 \cite{Planck_2020, Camphuis_2025, Louis_2025}, and CC constraints on $\dot{z}$. We used a nested sampling algorithm to sample the posterior probability and compute the Bayesian Evidence. The $f(Q)$ and phenomenological models are capable of solving the Hubble tension, having a mean posterior $H_0 \approx 72-74$~km/s/Mpc and 74~km/s/Mpc, respectively. The $f(Q)$ models assisted with the cosmological constant have $H_0 \approx 71$ km/s/Mpc, alleviating but not completely solving the Hubble tension. The reduced $H_0$ when more flexibility is added to the $f(Q)$ models arises because they struggle to solve the BAO anomaly. They have MDPS tensions of $7.55\sigma$, $2.65\sigma$, and $4.82\sigma$ C.L. for the Log, Exp, and Tanh models, respectively, when considering the CMB alone as a constraint on the model parameters (Table~\ref{tab:ChiSquaredResults_BAO_CMB}). The $\Lambda$CDM model shows a $2.65\sigma$ C.L. MDPS tension from the $(\Omega_m, H_0 r_d)$ parameter space, which is the conventional measure of the BAO anomaly. This tension reduces to $1.66\sigma$ C.L. in the DESI data space. The higher parameter space tension is indicative of a genuine anomaly, as discussed in Appendix~\ref{BAO_toy_model}. An interesting detail is that the Exp model slightly reduces the BAO anomaly in the $(\Omega_m, H_0 r_d)$ parameter space -- but not in data space.

The hybrid $f(Q)+\Lambda$ models have much reduced theoretical motivation because they are no longer alternatives to a cosmological constant. Moreover, they also show a $>2\sigma$ C.L. tension with the BAO data in both the data and parameter spaces, provided these models are constrained using CMB+$H_0$+CC (Table~\ref{tab:ChiSquaredResults_BAO_noDESI}). This is necessary because the models have extra flexibility at late times, making the model parameters difficult to constrain accurately using the CMB alone.

The models we consider predict an age of the Universe of $t_0 \approx 13.60-13.74$ Gyr, below the CMB + $\Lambda$CDM age of 13.805~Gyr \citep{Camphuis_2025} due to the high local $H_0$ measurements and the low observed comoving distance to $z \lesssim 1$ from DESI~DR2 BAO \citep{DESI_2025}. These slightly reduced ages are in mild $2\sigma - 3\sigma$ tension with the ages of the oldest Galactic stars and globular clusters, but the standard 13.8~Gyr prediction using CMB + $\Lambda$CDM is in $<2\sigma$ tension.

The Log model has a Bayesian Evidence much lower than $\Lambda$CDM, with $\ln B \approx -12$. The remaining models have $\ln B > 5$, so they all have very strong evidence over $\Lambda$CDM on the Jeffreys scale. The models with the highest Evidence are the phenomenological ones and $f(Q)$ Exp. The Phen models generally work well, but the Phen, PD model converges to $\Lambda$CDM too slowly to solve the Hubble tension. This is evident from the bottom panel of Figure~\ref{fig:DV_BAO}, where we find that $\alpha_\text{iso} \approx 0.97$ as $z \to 0$, indicating the model boosts $H_0$ by only about 3\%, whereas a 9\% boost is required \citep{H0DN_2025, Valentino_2025}. The Phen, exp and Phen, sech models work quite well overall, with MDPS BAO tensions of $2.13\sigma$ and $2.23\sigma$, respectively, both from data space. These tensions have no doubt been reduced by uncertainties in the model predictions due to the still uncertain parameters of models with two more degrees of freedom than $\Lambda$CDM (Figure~\ref{fig:phen_contours}). The best-fitting Phen models considering all datasets (including BAO) still depart systematically from the observed $\alpha_\text{iso}(z)$ (Figure~\ref{fig:DV_BAO}).

The Exp model is capable of providing a reasonable solution to the cosmological tensions studied in this paper. It solves the Hubble tension and slightly reduces the BAO anomaly in the $(\Omega_m, H_0 r_d)$ parameter space. The model is also preferred over $\Lambda$CDM using the Bayesian Evidence (+16.637). Furthermore, it is theoretically motivated as a dark energy candidate, in contrast to the models assisted with the cosmological constant and the phenomenological models discussed above. The $f(Q)$ models with $\Lambda$ typically have a higher Bayesian Evidence, showing that they fit the data better, even if their theoretical motivation is undermined. The exception is the Exp+$\Lambda$ model, which actually fits worse than Exp. This indicates that the Exp model does not require a cosmological constant term, allowing it to serve as an alternative to dark energy.

The phenomenological models offer a compelling \emph{empirical} solution to the tensions (especially in parameter space), but they are not theoretically motivated and predict a singularity in their dark energy EoS. This comes from the fluid energy density becoming negative and thus breaking the weak energy condition, implying that these models are unphysical in GR -- though they could potentially be valid solutions of another theory. The phenomenological models behave like $\Lambda$CDM across most of the relevant redshift range, with an extra expansion term at $z \leq 0.20$ to match the low-redshift data. This suggests that the solution to the Hubble and BAO tensions lies at low redshift, perhaps even just above the range of the distance ladder. 

Our exploration of late-time background solutions to the Hubble and BAO tensions shows that it is quite difficult to jointly solve both problems. While the Hubble tension can be solved by most of the considered models besides $\Lambda$CDM, these models run into difficulty with DESI~DR2, largely missing the observed points despite using the best-fitting parameters considering all datasets including DESI~DR2 (Figure~\ref{fig:DV_BAO}). Even the Exp model suffers from this issue: its best fit still overpredicts nearly all the isotropically averaged BAO distances reported by DESI~DR2. Given that these BAO measurements were anticipated by the local void solution to the Hubble tension without considering BAO datasets \citep{Haslbauer_2020, Banik_2025_BAO}, our results suggest that the Hubble and BAO tensions might be solved at the perturbative level.

\vspace{6pt} 






\section*{ ACKNOWLEDGEMENTS}

HD, JAN, and IB are supported by Royal Society University Research Fellowship 211046.

\section*{Data Availability}

Our code is publicly available \href{https://github.com/antonionajeraq/hubble_bao_tensions_fQ_gravity.git}{\faGithub}.

\appendix

\section{Comparison of best-fitting models to BAO measurements over the last 20 years}
\label{sec:FiguresBAO}

\begin{figure}
\centering
\includegraphics[width=13.0cm]{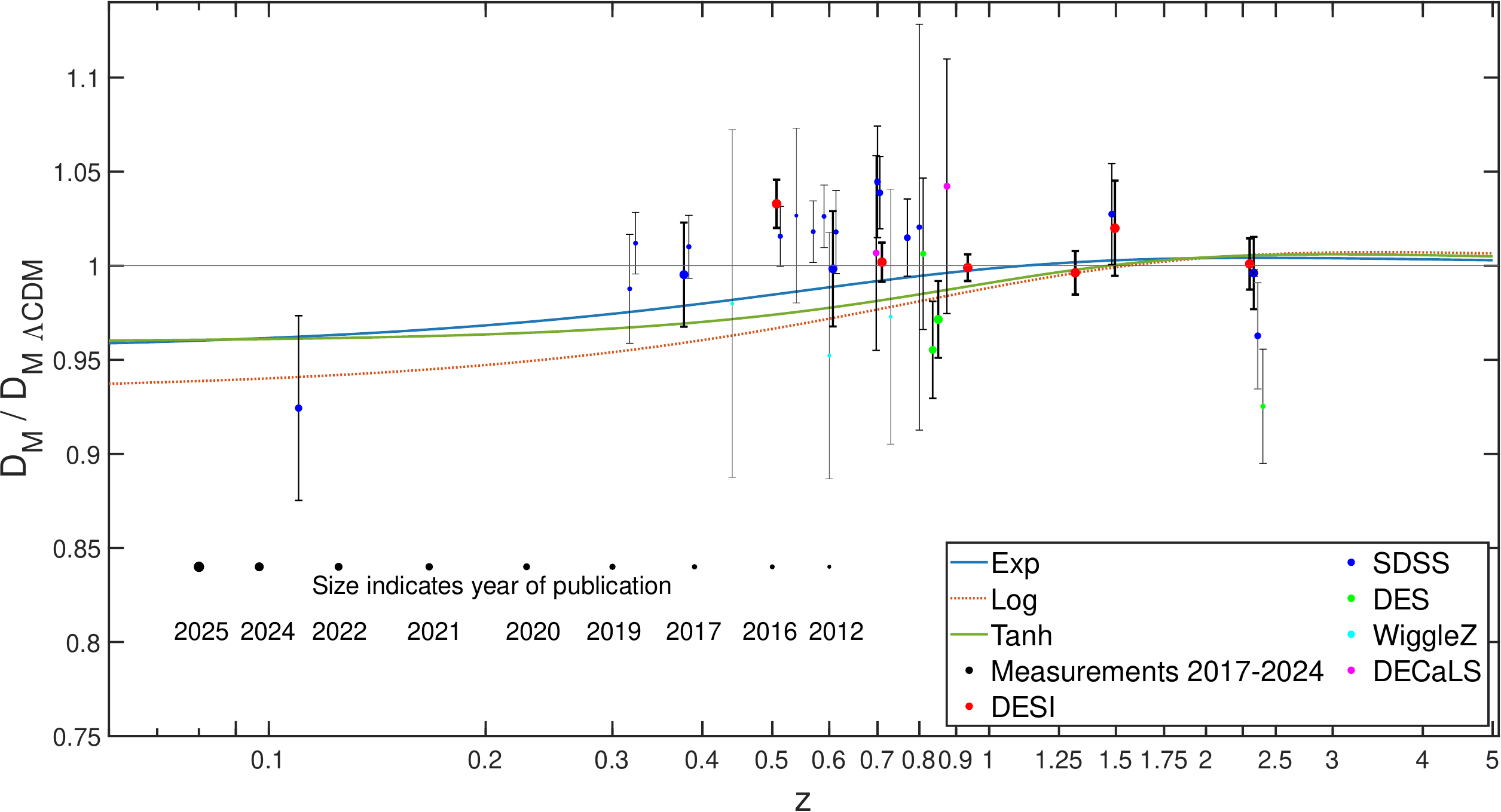} \\
\includegraphics[width=13.0cm]{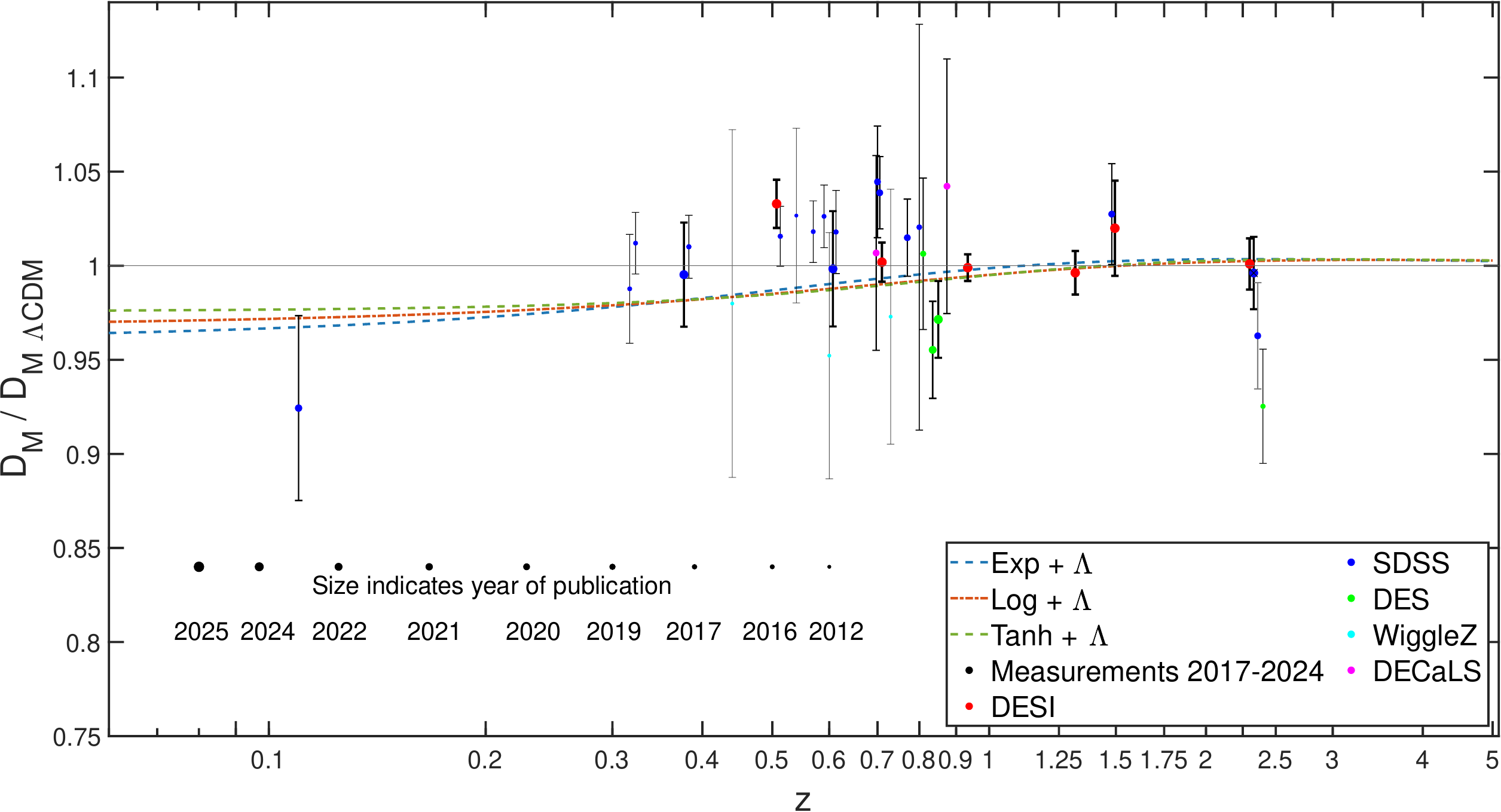}\\
\includegraphics[width=13.0cm]{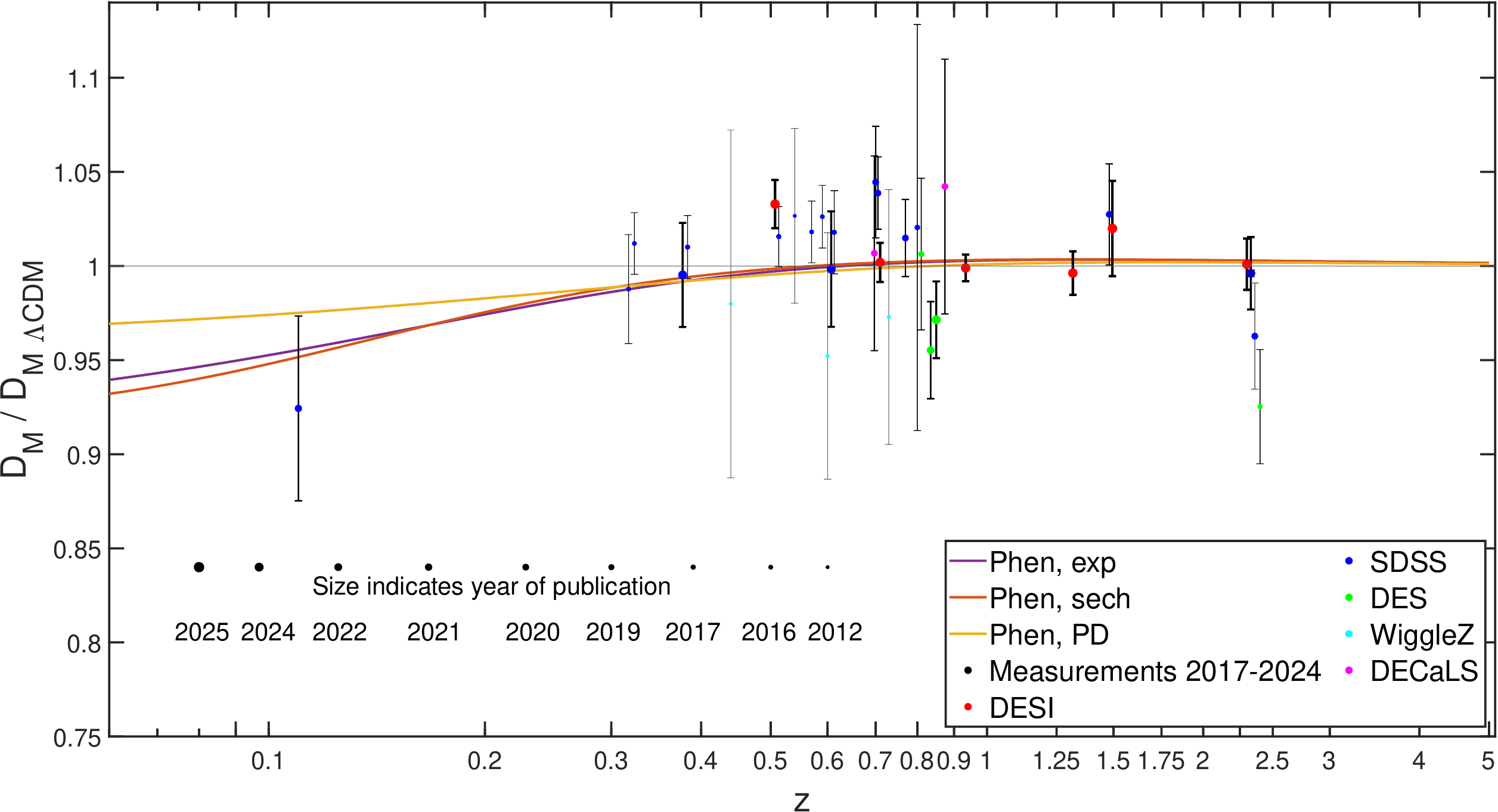}
\caption{Ratio between the comoving distance $D_M$ and the $\Lambda$CDM case. 
We take the best fits for all models with respect to the full data set ($H_0$+\emph{Planck}~2018+SPT+ACT+CC+DESI~DR2 BAO), including the $\Lambda$CDM case. We also plot the BAO data from the last 20 years for comparison. We present the results for \textbf{(a)} the $f(Q)$ models, \textbf{(b)} the $f(Q)$ models assisted with the cosmological constant, and \textbf{(c)} the phenomenological models. \label{fig:DM_BAO}}
\end{figure}

\begin{figure}
\centering
\includegraphics[width=13.0cm]{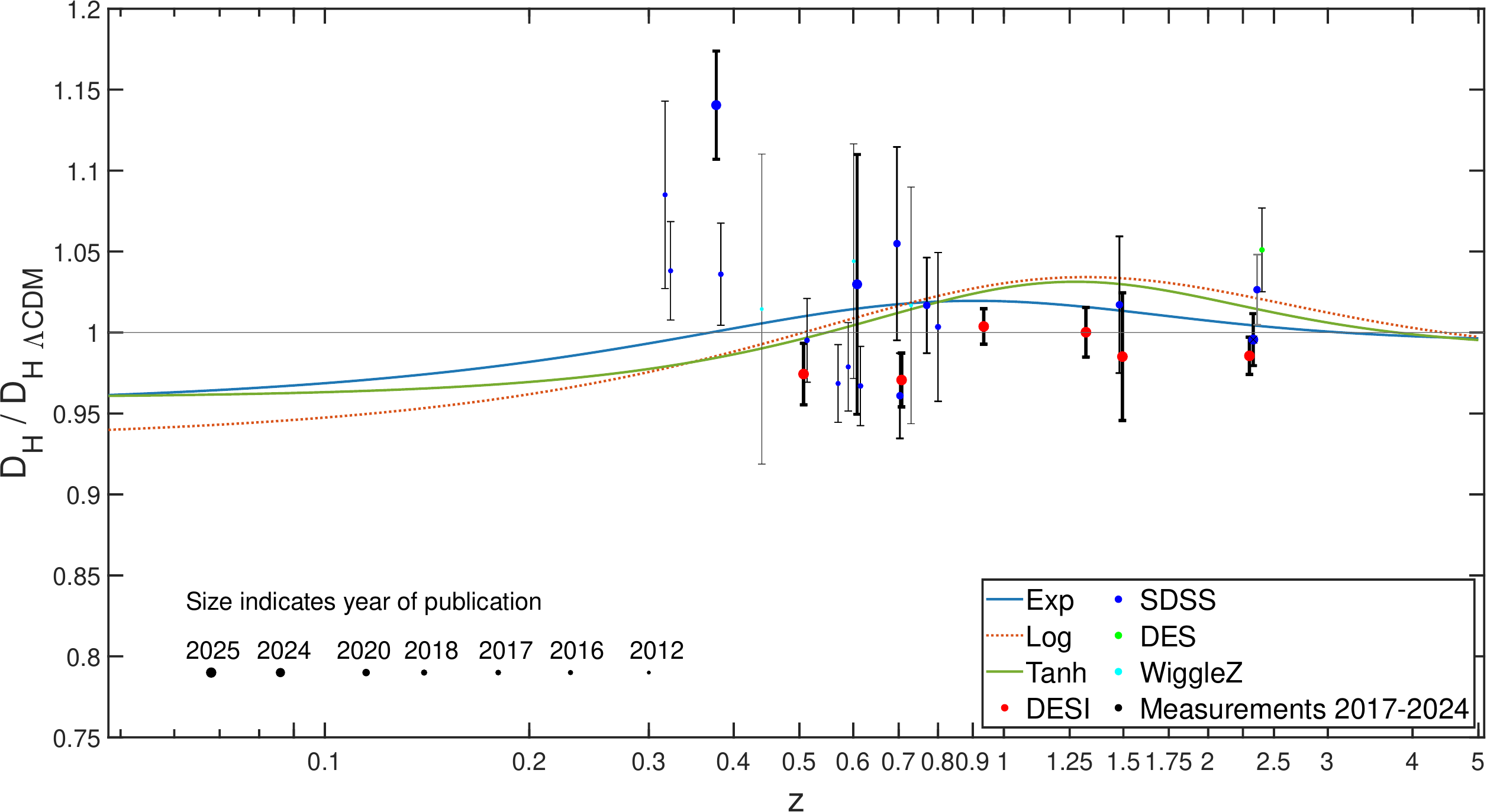} \\
\includegraphics[width=13.0cm]{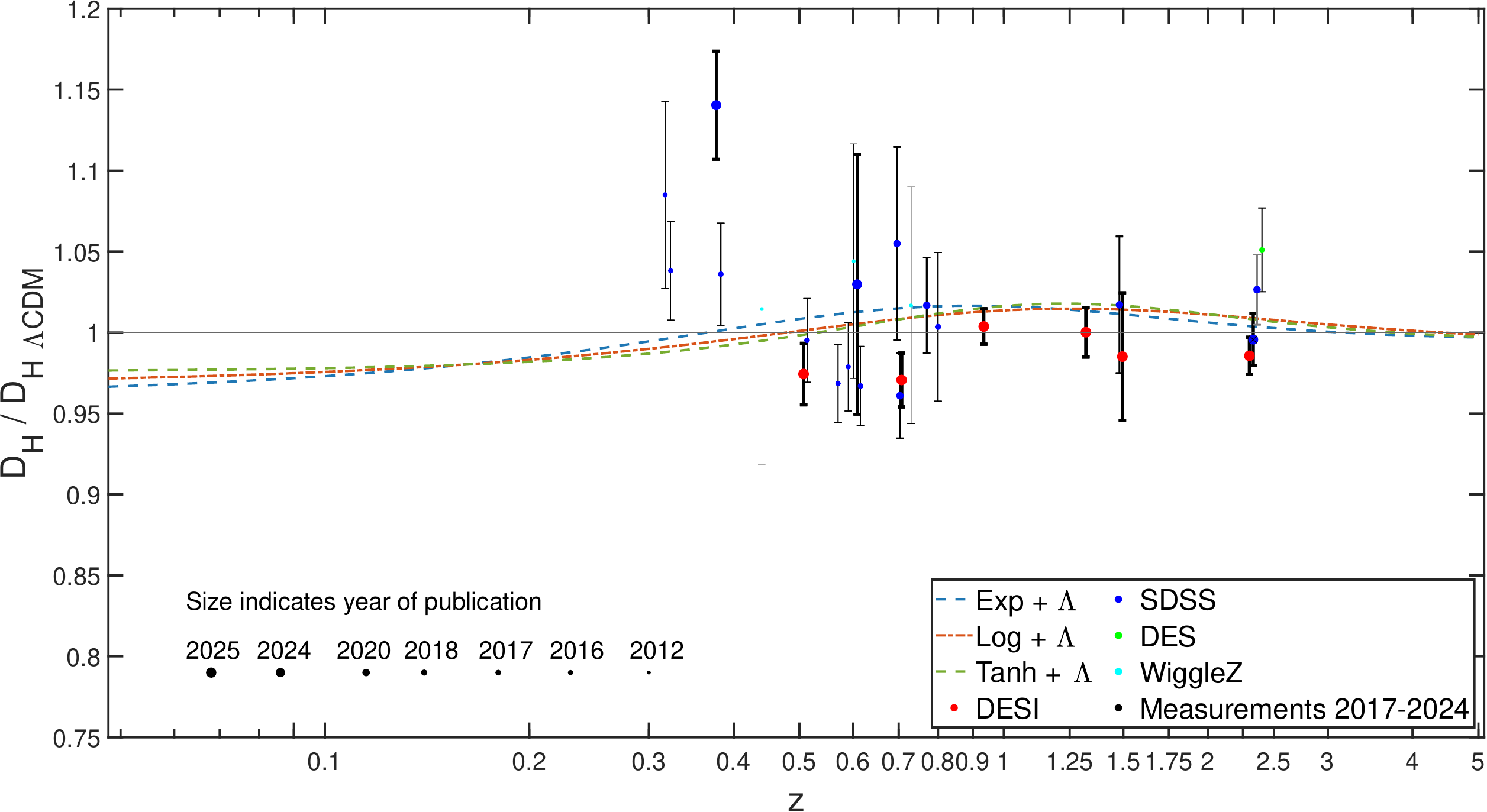}\\
\includegraphics[width=13.0cm]{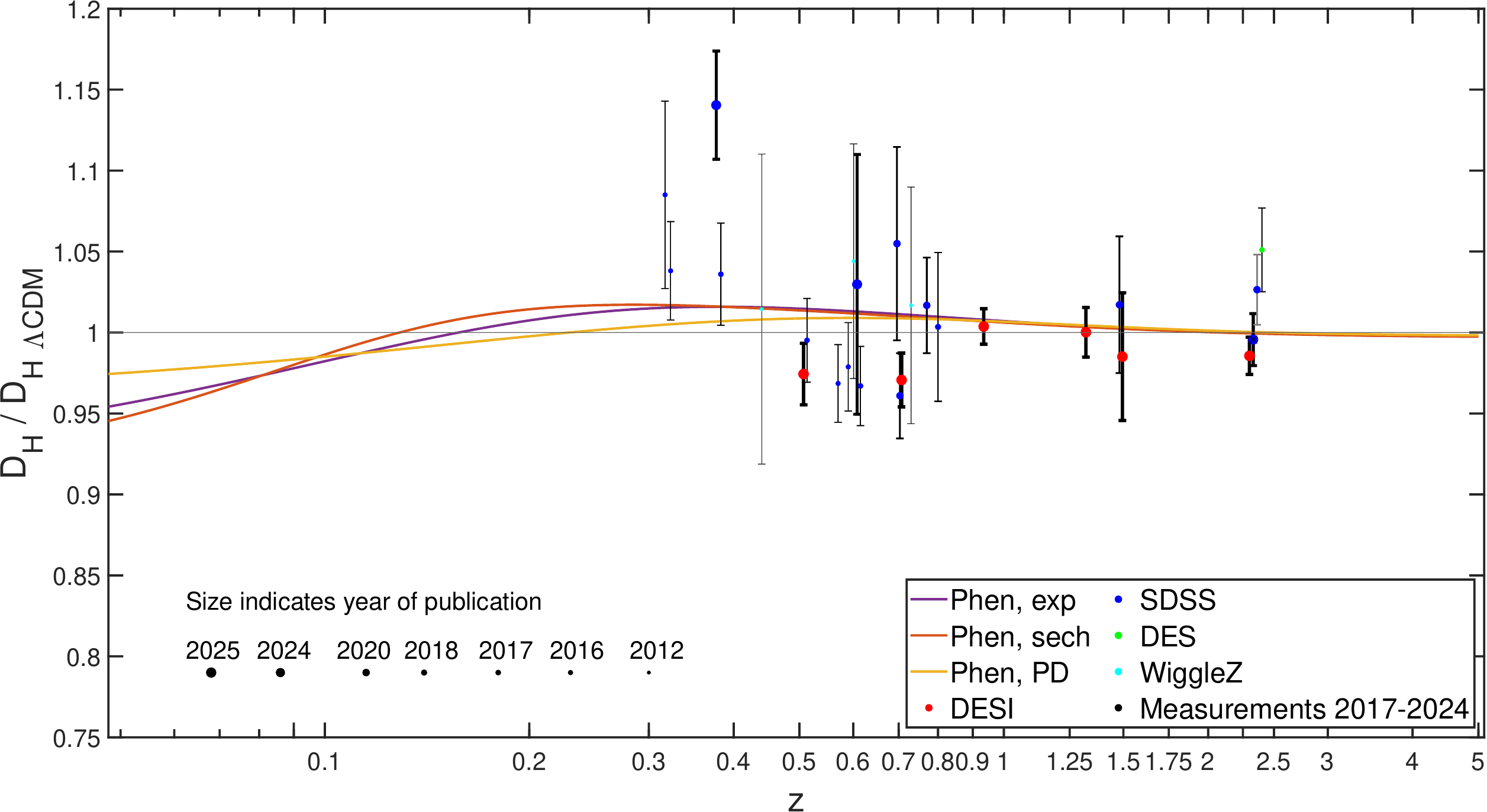}
\caption{Similar to Figure~\ref{fig:DM_BAO}, but showing instead $D_H$.}
\label{fig:DH_BAO}
\end{figure}

In our main analysis, we constrain our models using BAO measurements only from DESI~DR2 \citep{DESI_2025}. These are not the only BAO measurements. We therefore compare the best-fitting model in each case to all available BAO measurements over the last 20 years \citep[table 1 of][and references therein]{Banik_2025_BAO}. Figure~\ref{fig:DM_BAO} shows our results for $D_M$, while Figure~\ref{fig:DH_BAO} shows $D_H$ instead.

\section{CMB Likelihood}
\label{sec:CMB-detailed}


We compare our models to the CMB using the compressed CMB likelihood approach proposed in \cite{Lemos_2023, DESI_2025}, with correlated likelihoods for the parameters $(100 \, \theta_\star, w_b, w_{bc})$. This approach is valid as long as the model does not change early time physics, which is the case for all the models considered in this paper.

For the CMB likelihood, since we need to integrate up to the recombination redshift of $z_\star \approx 1090$, we must consider the photon and neutrino contributions. The photon density is
\begin{equation}
    \rho_\gamma = \frac{\pi^2 k^4}{15 \hbar^3 c^5} T_\text{CMB}^4,
\end{equation}
{where $k$ is Boltzmann's constant, $\hbar$ the reduced Planck constant, and $T_\text{CMB}$ the temperature of the CMB. We consider the mean result of the CMB temperature determinations ($T_\text{CMB} = 2.72548 \, \pm \, 0.00057 \, \text{K}$) and neglect the uncertainty \cite{Fixsen_2009}. The photon number density follows from standard blackbody physics.
\begin{equation}
    n_\gamma = \frac{2 \, \zeta(3)}{\pi^2} \left(\frac{k T_\text{CMB}}{\hbar c} \right)^3,
\end{equation}
where $\zeta(3) = 1.202$ is the Riemann zeta function with argument 3.

Neutrinos behave like photons at high redshift. The ultra-relativistic neutrino density
\begin{equation}
    \rho_{\nu, ur} \equiv \rho_\gamma R_{\nu/ \gamma},
\end{equation}
which is the neutrino energy density if they were massless. The ratio $R_{\nu/ \gamma}$ between neutrino and photon energy densities is applicable only in the early Universe, when it is given by
\begin{equation}
    R_{\nu/ \gamma} = \frac{7}{8} \left( \frac{4}{11} \right)^{4/3} N_\text{eff},
    \label{eq:R_nu_gamma}
\end{equation}
where $N_\text{eff} = 3.044$ is the effective number of neutrino species \citep{Akita_2020, Froustey_2020, Bennett_2021}.

The neutrino density in species $i$ (electron, muon, or tau) in the non-relativistic regime is given by
\begin{equation}
    \rho_{\nu, i} c^2 = n_{\nu, i} m_{\nu, i},
\end{equation}
where $m_{\nu, i}$ is the neutrino mass of flavour $i$ and $n_{\nu, i}$ is the number density of such neutrinos, which is given by
\begin{equation}
    n_{\nu, i} = \frac{3}{11} n_\gamma \left( \frac{N_\text{eff}}{3} \right)^{0.68}.
\end{equation}
If the neutrinos were thermalised, the exponent would be $3/4$. We use instead an exponent of 0.68 to ensure that in the non-relativistic limit, $w_\nu = \sum_i m_{\nu, i} c^2/(93.14 \, \text{eV})$ \citep{Lesgourgues_2012}. The lower exponent arises because the energy from electron-positron annihilation mostly goes into the high-energy tail of the neutrino energy distribution, which does not have enough time to thermalise before neutrino decoupling.

Unlike photons which are always relativistic, neutrinos become non-relativistic at some cosmic scale factor
\begin{equation}
    a_{nr} = \frac{1}{\sqrt{\left( \left. \dfrac{\rho_\nu}{\rho_{\nu, ur}} \right|_{z = 0} \right)^2 - 1}}.
    \label{eqn:a_nr}
\end{equation}
We use the following form for the transition between relativistic and non-relativistic behaviour:
\begin{equation}
    \rho_\nu = \left. \rho_{\nu,ur} \right|_{z = 0} (1+z)^4 \sqrt{1+\dfrac{1}{a^2_{nr}(1+z)^2}}.
\end{equation}
Note that the present neutrino density $\rho_\nu \gg \rho_{\nu,ur}$, what the neutrino density would have been if neutrinos were massless. We assume for simplicity that all the neutrino species have the same mass, since details of the brief relativistic to non-relativistic transition are unimportant -- only the neutrino mass sum of $\sum m_{\nu, i} = 0.06$~eV/$c^2$ is relevant to cosmology given present levels of precision. Thus, $a_{nr}$ is taken to be the same for all neutrino species.

We add the photon and neutrino contributions to the right-hand side of the Friedmann equations, with the neutrinos transitioning from ultra-relativistic to non-relativistic at $a = a_{nr}$. We now have the required ingredients to compute the comoving distance to any redshift, including that of recombination.

The observable $\theta_\star$ also depends on $r_\star$, the sound horizon at recombination. We compute $r_\star$ using the exact analytic result from \cite{Hu_1996}:
\begin{equation}
    r_\star = \frac{2c}{3} \sqrt{\dfrac{3 \, a_{eq}}{\Omega_{bc} H_0^2 \mathcal{R}_{eq}}} \ln \left( \dfrac{\sqrt{1+\mathcal{R}_\star}+\sqrt{\mathcal{R}_\star+\mathcal{R}_{eq}}}{1+\sqrt{\mathcal{R}_{eq}}} \right),
\end{equation}
where $\star$ subscripts denote recombination, eq subscripts denote matter-radiation equality (treating neutrinos as radiation; Eq.~\ref{eq:Friedmann_with_Lambda}), $\Omega_{bc} \equiv \Omega_b + \Omega_c$ is the density parameter for baryons + CDM, and $\mathcal{R} = 3\rho_b/(4\rho_\gamma)$ is related to the sound speed.

While $a_{eq}$ is easily determined from the cosmological parameters, estimating the recombination redshift $z_\star$ is more complicated. We use a modified version of the fitting formula from \cite{Hu_1996}:
\begin{eqnarray}
    \label{eqn:z_rec}
    z_\star &=& 1045 (1 + 0.00124 \, w_b^{-0.738}) (1+g_1 w_{bc}^{g_2}), \\
    g_1 &=& \frac{0.0783 \, w_b^{-0.238}}{1 + 39.5 \, w_b^{0.763}}, \\
    g_2 &=& \frac{0.560}{1 + 21.1 \, w_b^{1.81}}.
\end{eqnarray}
We found that using 1045 instead of 1048 brings $z_\star$ into much closer agreement with modern studies, presumably due to updates to the atomic physics constants entering into the calculation of precisely when recombination occurred. This modest change reduces $z_\star$ from 1092 to 1089, which is much closer to the value reported by the \emph{Planck} Collaboration \citep{Planck_2020}. It also means that when inferring $H_0$ and its uncertainty from the CMB alone, our result is in much closer agreement with the value reported by \cite{Camphuis_2025}. This is also the case when using $\mu_\text{obs}(100 \, \theta_\star, w_b, w_{bc})$ and its covariance as reported by the DESI Collaboration \citep{DESI_2025}, whose estimate of $H_0$ and its uncertainty based on the CMB alone now agrees much more closely with ours. Another argument for our change from $1048 \to 1045$ is that this slightly increases the ratio $r_d/r_\star$, bringing it into almost perfect agreement with the 1.0184 reported by \cite{Vagnozzi_2023}. However, we could not get reliable results simply by scaling our modern formula for $r_d$ (Eq.~\ref{eqn:rd}) by this factor to obtain $r_\star$, i.e., by assuming a fixed $r_d/r_\star$ ratio. Therefore, the best approach still seems to be that of \cite{Hu_1996}, but with a slightly modified $z_\star$ to better line up with modern results.

We consider the $\Lambda$CDM-based \emph{Planck}~2018 + SPT-3G + ACT~DR6 \cite{Planck_2020, Camphuis_2025, Louis_2025} MCMC chains from \footnote{\url{https://lambda.gsfc.nasa.gov/product/spt/spt3g_d1_bandp_liklyhood_get.html}} \cite{Camphuis_2025}. We extract the chains for the parameter vector $(100 \, \theta_\star, w_b, w_{bc})$, compute the maximum auto-correlation time $\tau_f$ of the three parameter chains, and apply a burn-in by removing the first $3 \, \tau_f$ steps. We then derive the mean values and covariance matrix, which are as follows:
\begin{eqnarray}
    \mu_\text{obs}(100 \, \theta_\star, w_b, w_{bc})^T &=& (1.04161, \, 0.02238, \, 0.14247), \\
    \mathbf{C}_\text{CMB} &=& 10^{-9} \times \begin{pmatrix}
        54.3960, & 0.806981, & -20.6489 \\
        0.806981, & 8.54260, & -13.2648 \\
        -20.6489, & -13.2648, & 692.160
    \end{pmatrix}.
\end{eqnarray}
We use these values in Eq.~\ref{eqn:CMB-likelihood}. We verified that when applying a similar procedure to the parameter $H_0$, we recover a value of $67.24 \pm 0.35$~km/s/Mpc from the MCMC chains, matching that stated in \cite{Camphuis_2025}. However, we cannot use their inference of $H_0$ from the CMB assuming $\Lambda$CDM at all times because our models introduce new physics at late times, altering the comoving distance to recombination.

\section{Toy model of the BAO anomaly}
\label{BAO_toy_model}

We have seen that when calibrating the $\Lambda$CDM model parameters using the CMB alone, the tension with DESI~DR2 BAO measurements is $1.66\sigma$ in data space but $2.65\sigma$ in parameter space (Table~\ref{tab:ChiSquaredResults_BAO_CMB}). To understand if this can plausibly arise by chance, we set up a toy model with points $(x_i, y_i)$ labelled by $i = 1-13$. The $x_i$ are analogous to redshifts, while the $y_i$ are analogous to $\alpha_\text{iso}$ or any other BAO-derived $\alpha$. We assume the $x_i$ have no uncertainty and uniformly cover the range $(-1, 1)$. The $y_i$ are assumed to have an uncertainty of $\sigma = 1$ in all cases. The latent values are drawn from two models:
\begin{enumerate}
    \item An `unbiased' model where $y_i = 0 \forall i$; and
    \item A `biased' linear model where $y_i = g x_i$, with fixed $g > 0$.
\end{enumerate}
To mimic observational errors, we add a Gaussian random number to each latent $y_i$.

The $\chi^2$ in data space is just
\begin{eqnarray}
    \chi^2_{\mathrm{DS}} ~=~ \sum_i y_i^2 \, .
\end{eqnarray}
We expect that on average, this is
\begin{eqnarray}
    \langle \chi^2_{\mathrm{DS}} \rangle ~=~ N \left[ 1 + g^2 \langle x^2 \rangle \right] \, ,
    \label{Analytic_chisq_DS}
\end{eqnarray}
where $N = 13$ is the number of data points and the average of the $x_i^2$ is $\langle x^2 \rangle = 0.389$. We convert $\chi^2_{\mathrm{DS}}$ into a data space tension by working out the likelihood of a higher $\chi^2$ for $N$ d.o.f. This is converted into the usual Gaussian equivalent tension for a single variable using the inverse survival function.

The parameter space we consider here is the slope and intercept of a line fit to the data, mimicking the $(\Omega_m, H_0 r_d)$ parametrisation of the BAO data. Bearing in mind that errors on the slope and intercept are uncorrelated when $\langle x \rangle = 0$, the $\chi^2$ in parameter space is
\begin{eqnarray}
    \chi^2_{\mathrm{PS}} ~=~ N \langle x^2 \rangle g_{\mathrm{obs}}^2 + N \overline{y}^2 \, ,
\end{eqnarray}
where the observationally inferred intercept is $\overline{y}$, the mean of the $y_i$, and the observationally inferred gradient is
\begin{eqnarray}
    \quad g_{\mathrm{obs}} ~=~ \frac{\sum_i x_i y_i}{\sum_i x_i x_i} \, .
    \label{eq:g_inferred}
\end{eqnarray}
We expect that the average $\chi^2_{\mathrm{PS}}$ is
\begin{eqnarray}
    \langle \chi^2_{\mathrm{PS}} \rangle ~=~ 2 + N g^2 \langle x^2 \rangle \, .
    \label{Analytic_chisq_PS}
\end{eqnarray}
This is because on average, each of the two parameters should contribute 1 even if $g = 0$, but in general there is an additional contribution because we expect $g_{\mathrm{obs}}$ to be centred on the true $g$. The parameter space tension is quantified by finding the likelihood that $\chi^2 \geq \chi^2_{\mathrm{PS}}$ for 2 d.o.f.

\begin{figure}
    \centering
    \includegraphics[width=9.0cm]{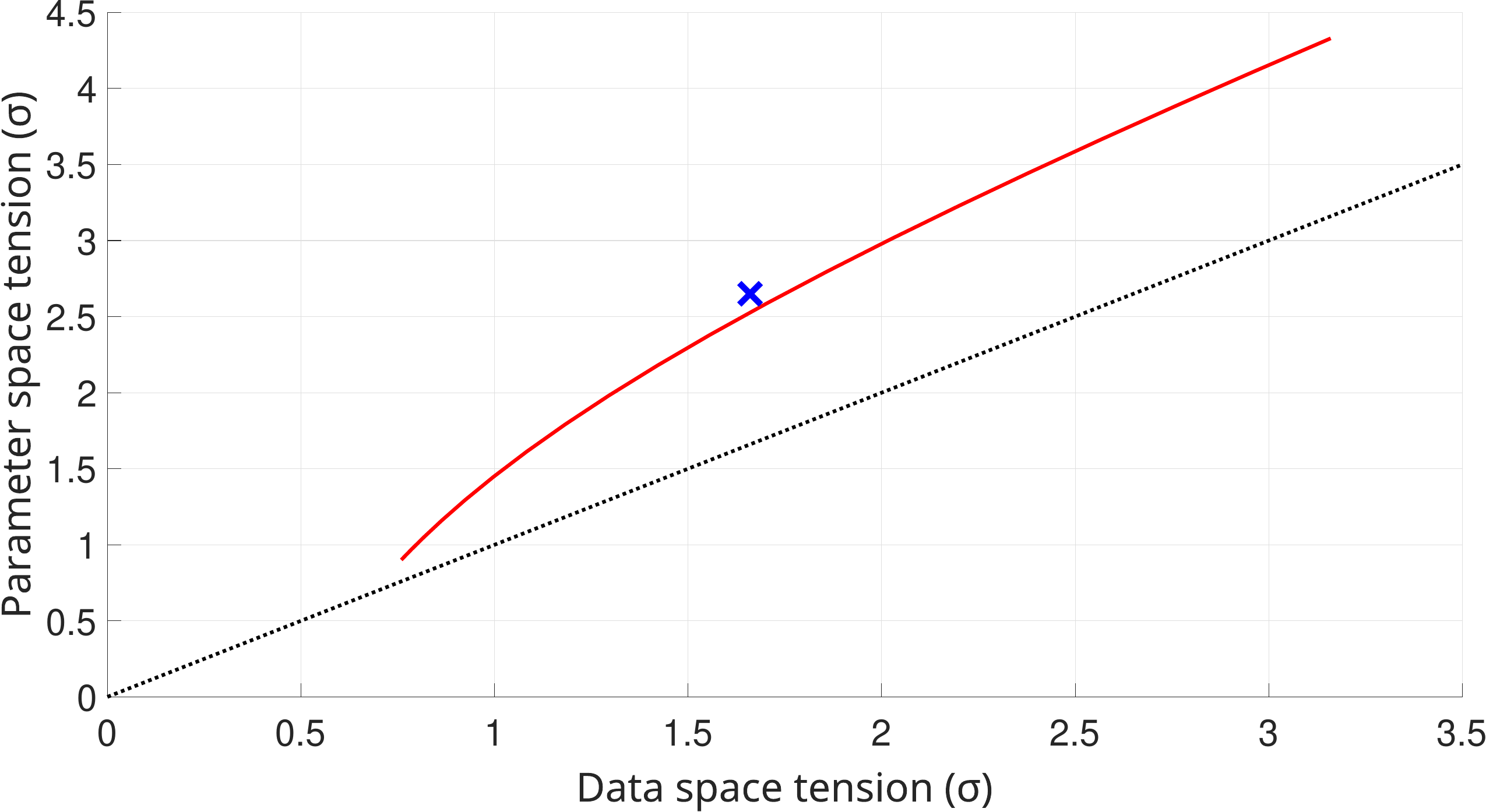}
    \caption{Analytic estimates of the data and parameter space tensions in toy models where the latent $y_i = gx_i$ for $g = 0-2$ (solid red line), with $\chi^2$ given by Equations~\ref{Analytic_chisq_DS} and \ref{Analytic_chisq_PS}, respectively. These are with respect to the `prediction' that $y = 0$. Notice that the expected result lies above the dotted black line of equality, indicating more parameter space tension. The blue cross shows the significance of the BAO anomaly in $\Lambda$CDM calibrated using CMB alone (Table~\ref{tab:ChiSquaredResults_BAO_CMB}).}
    \label{fig:Analytic_biased_tensions}
\end{figure}

Figure~\ref{fig:Analytic_biased_tensions} shows our analytic estimate of the data and parameter space tensions in our toy model for $g = 0-2$. The solid red line shows that the parameter space tension is larger because the inferred slope (Eq.~\ref{eq:g_inferred}) is ideally suited to capture the true nature of deviations from the comparison model ($y = 0$). The blue cross shows the BAO-CMB tension in $\Lambda$CDM (Table~\ref{tab:ChiSquaredResults_BAO_CMB}). Remarkably, this lies almost exactly on the expected track for our biased toy model. This suggests the data and parameter space significance levels of the BAO anomaly can be readily understood as a systematic departure from the $\Lambda$CDM expectations.

The above argument implicitly assumes that the real BAO anomaly can be efficiently captured in the $(\Omega_m, H_0 r_d)$ parameter space. This is definitely the case for DESI~DR2: our results in Table~\ref{tab:ChiSquaredResults_BAO_noDESI} show that the observations can be fit extremely well in $\Lambda$CDM using \emph{some} values of $(\Omega_m, H_0 r_d)$, even though these values are in tension with fits to the CMB alone. Roughly speaking, altering $H_0 r_d$ is analogous to changing the intercept in our toy model, while altering $\Omega_m$ is analogous to changing the slope. The real BAO anomaly can be thought of as an extra tilt in the space of $\alpha$ vs $z$, where $\Lambda$CDM by definition predicts $\alpha = 1 \forall z$. As a result, the standard $(\Omega_m, H_0 r_d)$ parametrisation is sufficient to capture the BAO anomaly for the time being. In general, new physics might of course be impossible to efficiently capture in a conventional parameter space. For instance, we can easily envisage that in our toy model, the true $y_i = x_i^2 - 1/3$, which does not create any offset in either the slope or the intercept. This would cause the parameter space tension to remain small, but the U-shaped residuals would eventually show up in data space. These arguments highlight the importance of considering the tension in both spaces, which we summarise using the MDPS tension. 

\begin{figure}
    \centering
    \includegraphics[width=8.5cm]{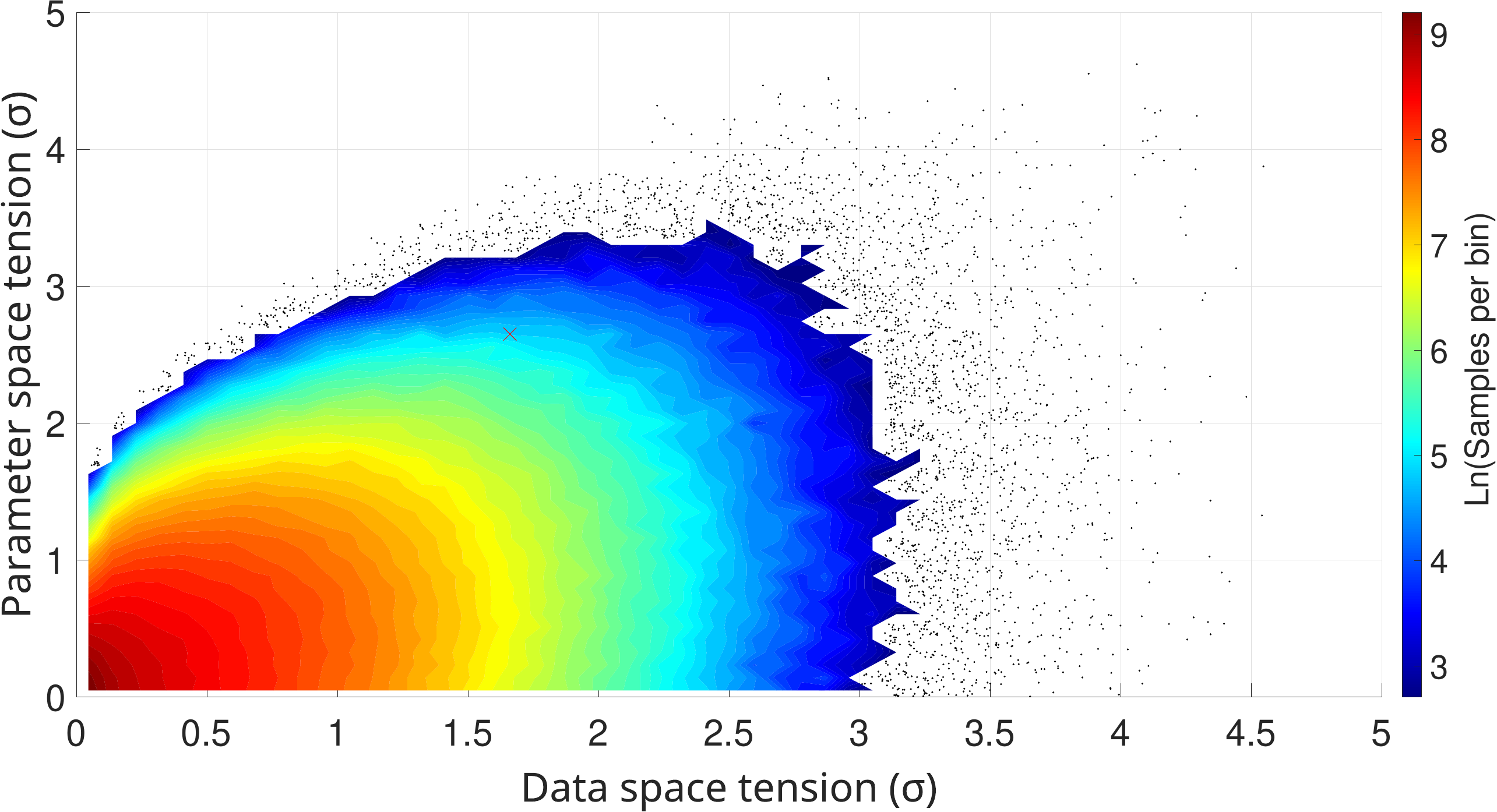}
    \hfill
    \includegraphics[width=8.5cm]{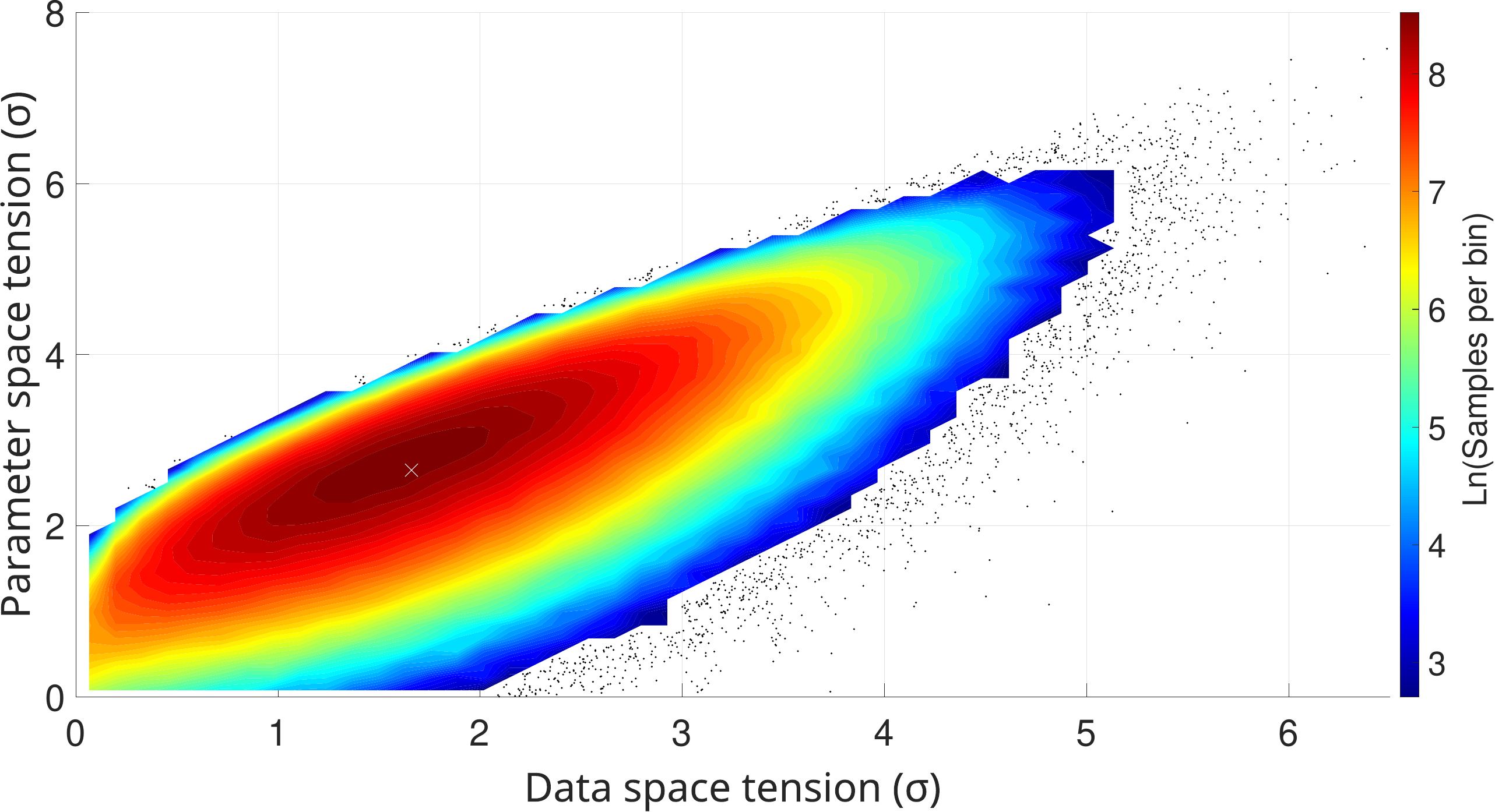}\\
    \caption{Heat maps showing the distribution of parameter and data space tensions in mock datasets where the latent $y_i = 0$ (\emph{left panel}) or $1.3 \, x_i$ (\emph{right panel}), the comparison model being $y = 0$. The black dots show individual Monte Carlo realisations in low density regions, while higher density regions are shown using a logarithmic heat map. The cross in each panel shows the significance of the BAO anomaly in $\Lambda$CDM calibrated using CMB alone (Table~\ref{tab:ChiSquaredResults_BAO_CMB}). Notice that this is a significant outlier from the distribution in the unbiased model, but almost at the highest density point in the biased model.}
    \label{fig:Toy_model_contours}
\end{figure}

To understand whether the BAO anomaly can arise by chance, we set up $10^6$ Monte Carlo trials of our unbiased toy model. The left panel of Figure~\ref{fig:Toy_model_contours} shows the resulting distribution of data and parameter space tensions. The BAO-CMB tension is indicated using the red cross. It is immediately apparent that this is an outlier. Moreover, even if there is a $1.66\sigma$ data space tension, the parameter space tension is typically not as high as $2.65\sigma$. We can quantify the combined significance as the proportion of mock datasets in our toy model which lie outside the contour through the actual BAO-CMB tension. The result is 2.48\%, corresponding to a $2.24\sigma$ tension.


The right panel of Figure~\ref{fig:Toy_model_contours} shows the corresponding results with our biased toy model, where we set $g = 1.3$ to approximately match the $1.66\sigma$ data space BAO-CMB tension. The white cross shows the observed BAO-CMB tension. This lies almost exactly at the most likely point in Monte Carlo trials of our biased toy model. As a result, 89.4\% of the mock datasets give a tension level outside the contour through this point. This corresponds to a tension of only $0.13\sigma$. The excellent agreement is to be expected on analytic grounds (Figure~\ref{fig:Analytic_biased_tensions}).

Although we have considered a 2D parameter space, our biased toy models have an offset only in the gradient. This suggests a higher tension could be achieved if focusing only on $g$. We do not take this approach because our toy model is meant to illustrate the standard $(\Omega_m, H_0 r_d)$ parametrisation of the BAO anomaly. A simpler 1D parametrisation is possible if we focus on a standard cosmological parameter, avoiding look elsewhere effects from creating statistics tailored to fine details of the BAO anomaly. One parameter that has been considered is $\sum m_{\nu, i}$. Working out its impact on cosmological observables for different values allows an extrapolation to negative $\sum m_{\nu, i}$, which has a clear phenomenological meaning in this context. We might expect that since this largely tilts the relation between $\alpha$ and $z$ due to neutrino rest mass only becoming an important consideration at low $z$, the mostly low-redshift BAO anomaly might give an even higher parameter space tension when focusing on $\sum m_{\nu, i}$ rather than on $(\Omega_m, H_0 r_d)$, which allows both a tilt and a change in normalisation. Indeed, $\sum m_{\nu, i}$ is pushed towards unphysical negative values, conflicting with the lower limit from terrestrial neutrino oscillation experiments at $3\sigma$ C.L. \citep{DESI_2025_neutrinos}. This neutrino mass anomaly indicates a problem with $\Lambda$CDM at low redshift.

In summary, a genuine BAO anomaly might well be expected to have a smaller significance in data space because this is a blind test of deviations from the model predictions. If there are in fact systematic deviations resembling a change in one or both parameters, the deviations are better captured by the parameter space tension, which should be higher. This is indeed the case (Figure~\ref{fig:Analytic_biased_tensions}). However, the parameter space tension can give an incomplete picture because deviations with a different shape might not show up here, e.g. large parabolic residuals might leave the slope and intercept unaltered. In this case, the data space tension would be large, though the lack of any `template' for the anomaly would make it more difficult to discern from the null hypothesis. This shows why it is important to use the MDPS tension.

Our toy model highlights that although the BAO anomaly is not yet firmly established because it could in principle be due to a rare statistical fluctuation, the higher parameter space tension is much more readily understood as a systematic departure from $\Lambda$CDM expectations at low $z$ (Figure~\ref{fig:Toy_model_contours}). This is reinforced by the even higher tension in the neutrino mass sum, which has an effect only at low redshift.



\bibliographystyle{unsrt2authabbrvpp.bst}

\bibliography{fQ_bbl}

\end{document}